\documentclass[traditabstract]{aa}

\usepackage[dvips]{graphics}
\usepackage{natbib}
\usepackage{graphicx}
\usepackage{amsmath}
\usepackage{xcolor}

\bibpunct{(}{)}{;}{a}{}{,} 
\newcommand{\ind}[1]{_{\rm{#1}}} 
\newcommand{\xir}{\xi\ind{r}}
\newcommand{\xih}{\xi\ind{h}}
\newcommand{\Icore}{I\ind{core}}
\newcommand{\Teff}{T\ind{eff}}
\newcommand{\Omegacore}{\langle\Omega\rangle\ind{core}}
\newcommand{\numax}{\nu\ind{max}}
\newcommand{\nmax}{n\ind{max}}
\newcommand{\Dnu}{\Delta \nu}

\begin{document}

\title{Seismic diagnostics for transport of angular momentum in stars}
\subtitle{2. Interpreting observed rotational splittings of slowly-rotating red giant stars}

\author{M.J. Goupil\inst{1}
\and
B. Mosser\inst{1} 
\and
J.P. Marques\inst{2,1}
\and
R.M. Ouazzani\inst{3,1}
\and
K. Belkacem\inst{1}
\and
Y. Lebreton\inst{4}
\and
R. Samadi\inst{1}
}

\institute
{
LESIA, CNRS UMR 8109, Universit\'e Pierre et Marie Curie, Universit\'e Denis Diderot,
Observatoire de Paris, F-92195 Meudon, France;
\and
Georg-August-Universit\" at G\" ottingen, Institut f\" ur Astrophysik, Friedrich-Hund-Platz 1, D-37077 G\" ottingen, Germany
\and
Institut d'Astrophysique, G\' eophysique et Oc\' eanographie de l'Universit\' e de Li\` ege, All\' ee du 6 Ao\^ ut 17, 4000 Li\` ege, Belgium
\and
Observatoire de Paris, GEPI, CNRS UMR 8111, F-92195 Meudon, France
}

\abstract
{Asteroseismology with the space-borne missions CoRoT and \emph{Kepler} provides a powerful
mean of testing the modeling of transport processes in stars.
Rotational splittings are currently measured for a large number of red giant stars
and can provide stringent constraints on  the rotation profiles.

The aim of this paper is to obtain a theoretical framework for understanding the properties of the
observed rotational splittings of red giant stars with  slowly  rotating cores.
This allows us to establish  appropriate seismic diagnostics
for rotation of these evolved stars. 

Rotational splittings for stochastically excited dipolar modes are
computed adopting a first-order perturbative approach for two $1.3 M_\odot$
benchmark models assuming slowly rotating cores.

For red giant stars with slowly rotating cores, we show that the variation of the rotational splittings of $\ell=1$ modes
with frequency 
depends only on the large frequency separation,
the g-mode period spacing, and the ratio of the average envelope  to
  core rotation rates  (${\cal R}$). This leds us to propose a way to  infer directly
  ${\cal R}$ from the observations. This method is validated using the Kepler red giant
  star KIC 5356201. Finally, we provide a theoretical support for the use of a
  Lorentzian profile to measure the observed splittings for red giant stars.
 }
 

\maketitle

\section{Introduction}

Stellar rotation plays an important role  on the structure and evolution 
of stars. The problem of transport of angular momentum inside stars is not yet fully understood, however.
 Several mechanisms seem to be active although they are either only approximately described
or not modeled at all. 
In order to study the internal transport and evolution of angular momentum with time, 
one  needs  observational  constraints on physical quantities affected by 
such transport processes. In particular, the 
 knowledge of the internal rotation profiles and their evolution with time is crucial.
 An efficient way is  to obtain seismic information on the internal rotation profile of stars. 

 The ultra-high precision photometry (UHP) asteroseismic space missions such
 as CoRoT \citep{2006cosp...36.3749B} and {\it Kepler} \citep{2010Sci...327..977B} offer such an opportunity. 
Because cool low-mass stars have a convective envelope, 
oscillations can be stochastically excited as in the solar case. Red giant stars are of particular interest here. 
 Stochastically excited non-radial modes were detected with CoRoT \citep{2009Natur.459..398D}.
 Analyses of these data revealed the oscillation properties of a large number of 
 these stars 
 \citep[e.g.,][]{2009A&A...506..465H,2010A&A...517A..22M,2011A&A...532A..86M,2011Natur.471..608B}.
 Their frequency spectra show similarities as well as  differences with the  solar case
  \citep{2011A&A...525L...9M}. 
  Several works have investigated the properties of the  non-radial oscillation
   modes of red giant stars \citep[e.g.,][]{1975PASJ...27..237O,1977AcA....27...95D,
   2001MNRAS.328..601D,2009A&A...506...57D,2010ApJ...721L.182M}. 
  The internal properties of  a red giant star,  a dense core   and a diffuse envelope,  
  give rise to  both gravity-type oscillations  in the central region and 
 acoustic-type oscillations in the envelope. These oscillations were first called  mixed
 modes by \cite{1971AcA....21..289D} for a Cepheid model and \cite{1974A&A....36..107S} 
for condensed models representative 
 of a red giant structure. 
This has already been proved useful to probe the structure of these stars 
\citep[e.g., ][]{2011Natur.471..608B,2011A&A...532A..86M}. 

More recently,  
measurements of rotational splittings of red giant stars have been obtained 
by using the  {\it Kepler} observations 
\citep[e. g.,][]{2012ApJ...756...19D,2012Natur.481...55B,2012A&A...540A.143M}. 
These splittings provide the first direct insight into the rotation profiles of the 
innermost layers of stars \citep{2012ApJ...756...19D}.  
Observations  \citep{2012arXiv1209.3336M} reveal that a sub-sample  of 
oscillating red giants  have  complex frequency spectra. Their central layers are
 likely to rotate quite fast. A correct rotation rate ought then be deduced from  direct calculations of
 frequencies  in a non-pertubative approach (see the case for polytropes  and acoustic modes in 
\citealt{2006A&A...455..621R,2011LNP...832..259L,2012arXiv1209.5621O} 
 and  for g modes in \citealt{2010A&A...518A..30B,2011arXiv1109.6856B}).
This will be necessary in order to decipher the  complex spectra for fast rotation.  
On the other hand, the rest of the sample shows simple power spectra  that can 
be easily identified without ambiguity. These identifications then lead to quite small rotational
 splittings 
\citep{2012Natur.481...55B,2012arXiv1209.3336M}, thus to slowly rotating cores. 

On the theoretical side, the central rotations predicted theoretically by 
 standard models currently including rotationally induced mixing are 
 far too fast \citep[e.g., ][]{2012A&A...544L...4E} and cannot account for these
observed slow core rotation in red giants.

Motivated by these recent results, we have started a series of studies 
dedicated to an understanding of the rotational properties of oscillating  evolved stars. 
\cite{Marques2012} (hereafter Paper I) computed the rotation profiles and their evolution 
with time from the pre main-sequence to the red giant branch (RGB) for 1.3 $M_\odot$  stellar 
models with an evolutionary code where rotationally induced mixing had been implemented. 
The authors then
followed the evolution of the  rotational splittings calculated to  first-order 
 approximation in the rotation rate  using  
 the rotation  profiles predicted by their evolutionary computations.  
 The authors  found that for low-mass stars,  the first-order approximation
  is valid for most  stochastically excited oscillation modes
 for all evolutionary stages except the RGB.
 For red giants, the authors showed that there is indeed room enough in the uncertainties 
of the description of transport of angular momentum to decrease significantly the core rotation of red
giant models \citep[see also][]{Meynet2012}. However, they still predict  
a core  rotation that  is still too large
compared with the observed rotational splittings, meaning that  some additional processes 
are operating in the star to slow down its core. To test such possible additional mechanisms, 
investigations of the specific properties of the rotational splittings of red giants with slow core rotation 
must be carried out,  essentially because of the  particular dual nature of
the excited red giant modes.

We thus defer the study of red giants with fast rotating cores to the 
third paper of this series. 
We focus here  on red giants with slowly rotating cores. 
We present a study of the properties 
of  linear rotational splittings, computed for adiabatic oscillation modes, in the 
frequency range of observed modes in red giant stars.   
We assume arbitrarily slow rotation profiles and  then study essentially 
the impact of the properties of the modes  on the rotational splittings. 
  Our goal is to provide a theoretical framework for the interpretation of  the  observed
rotational splittings of red giants in terms of core and envelope rotation.

\begin{figure}[t]
\centerline{\includegraphics[height=7.5cm,width=9cm]{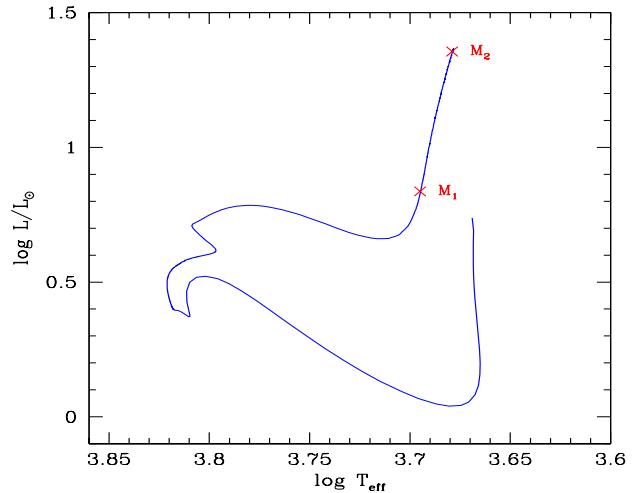}}
\caption{Hertzsprung-R\"ussell (HR) diagram showing a $1.3 M_\odot$ evolutionary
 track. The red  crosses indicate the selected models as described in 
Sect.~\ref{Sect:models} and in Table.~\ref{Tab1}.}
\label{HR}
\end{figure}

The paper is organized as follows; in Sect. \ref{structure}, we describe the properties of our models. 
In order to interpret the rotational splittings, we need first to identify the physical nature of 
the excited  modes. This is done in Sect. \ref{class} where we distinguish two
classes of mixed modes. 
 In Sect. \ref{splitting}, we compute    theoretical   rotational splittings using adiabatic eigenfrequencies
 and eigenfunctions obtained  
  for  $\ell$ $=1$ modes. 
 The studied  frequency  range corresponds to expected stochastically excited modes.
We then show that the rotational splittings exhibit repetitive patterns with frequency that 
can be folded into a single pattern as they carry nearly the same information.  We then consider in detail such pattern
for one of our models and investigate the
 information provided by the corresponding rotational splittings. 
 In Sect.~\ref{seismicconstraints},  we establish an approximate formulation 
 for the linear rotational splittings  in terms of the core contribution to mode inertia. The approximate
 expression   fits well the numerical ones and allows us their interpretation. 
 We  also give a procedure that enables ones to
derive the mean rotation of the core and of the envelope. Proceeding further in the approximations in 
Sect.\ref{interpretation},  we find an approximate formulation of the core contribution 
to mode inertia --  thus to rotational splittings --
that depends only on observable quantities aside rotation. 
We validate our theoretical results on the Kepler observations of the 
star KIC 5356201 (Beck et al. 2012). Finally, we end with some conclusions in Sect.\ref{conclusion}.

\begin{table}
\label{Tab1}
\caption{Fundamental parameters of the 1.3 $M_\odot$ stellar models.  }
\centering
\begin{tabular}{c c c c }
\hline\hline
model                          & M1           & M2    \\
\hline
$ R/R_\odot$                   & 3.58         &6.91       \\
$\log \Teff$ (K)               & 3.694        &  3.758       \\
$\log L/L_\odot$               & 0.825        &     1.345     \\
$\rho_c $ ($10^{7}$ kg/m$^3$)  &  7.06        &  14.05                     \\
$\rho_c/\bar{\rho}$ ($10^{6}$) &  5.12        & 75.85                      \\
$r_{CZ} (m_{CZ})$          &  0.25 (0.33) & (0.078)   (0.197)          \\
$r_\epsilon (m_\epsilon)$      &  0.01 (0.13) & ($4.5~10^{-3}$)   (0.168)  \\
$\Delta \nu$ ($\mu$Hz)         &  23.12       &  8.47                      \\
$\Delta \Pi$ (s) & 96.64  & 71.14 \\
$\numax$ ($\mu$Hz)  & 342.6  &  83.5   \\
$\nmax$&  14& 9 \\
\hline
\hline
\end{tabular}
\end{table}

\section{Structures of stellar models and rotation profiles}\label{structure} 
\label{Sect:models}

We consider $1.3 M_\odot$ models of red giants as this mass is typical for observed
red giant stars \citep{2010A&A...517A..22M}. These models (structure and rotation profile) 
have been evolved from PMS as described in Paper I.

 \begin{figure}[t]
\centerline{\includegraphics[height=8cm,width=9cm]{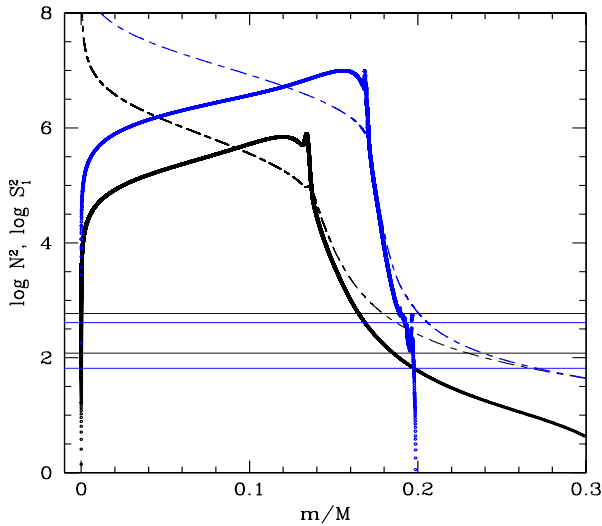}}
\caption{Logarithm of normalized  squared Brunt-V\"ais\"al\"a  $N^2$  (solid curve) and ($\ell=1$) 
Lamb  frequencies  $S_l^2$ (dash-dotted curve) as a function of  the normalized mass $m/M$ for
   model M1 ($\log_{10} \sigma_{\rm max}^2 =2.60$)  (black curves) and M2 
   ($\log_{10} \sigma_{\rm max}^2 =2.32$)  (blue curves). The thin horizontal solid 
   lines  indicate the frequency range for expected excited
  oscillation modes: black for model M1 and blue for M2.}
\label{vais1}
\end{figure}

\subsection{Selected  models} 

The first model (M1) lies at the base of the red giant 
branch (Fig.~\ref{HR}).  Its outer convective zone occupies 67\,\% in mass of the outer 
envelope and is located at $\Delta m/M=0.2$ above the H-shell burning. 
The second model (M2) is further up on the ascending branch (Fig.\ref{HR}). 
Its outer convective zone occupies 80 \% of the outer envelope and is at  
$\Delta m/M=0.015$ above the H-shell burning.  
 Table \ref{Tab1} lists the stellar 
 radius in solar unit, the effective temperature, luminosity in solar unit, 
 the central density $ \rho_c$, 
the ratio $ \rho_c/\bar{\rho}$  with $\bar{\rho}=3 M/(4\pi R^3) $, 
the radius at the base of the outer convective zone, the radius of maximum nuclear production rate, both
normalized to the stellar radius, 
the seismic quantities: 
 the large separation $\Dnu$, the g mode period spacing $\Delta \Pi$ (in s), 
 the frequency at maximum power spectrum intensity $\numax$, 
the radial order $\nmax$ at $\numax$.

The propagation diagrams  \citep{1975PASJ...27..237O}  for models M1 and M2 are  displayed in Fig.~\ref{vais1}. 
Only the central part is shown as it plays a major role in determining the properties of the 
oscillation modes of red giants (see Sect.~\ref{class}). The Brunt-V\"ais\"al\"a ($N$) and 
Lamb ($S_\ell$) frequencies take their usual definition, \emph{i.e.}, 
\begin{eqnarray}
N^2 &= &  \frac{g}{r} \left(\frac{1}{\Gamma_1} \frac{{\rm d} \ln p}{{\rm d}\ln r}- \frac{{\rm d}\ln \rho}{{\rm d}\ln r} \right) \, , \\
S_\ell^2 &= & \frac{\Lambda c_s^2}{r^2} \, ,
\end{eqnarray}
where $\Lambda = \ell(\ell+1)$, $\ell$ is the angular degree of the mode, $c_s^2$ is the 
squared sound speed,  the gravity $g=Gm/r^2$, and $p,\rho,\Gamma_1$ have their usual meaning. 
Unless otherwise stated, we consider all squared frequencies $N^2, S_\ell^2$ to be normalized to $GM/R^3$ 
and $\sigma^2 =\omega^2/(GM/R^3)$ (where $\omega$ is the mode pulsation in rad/s).

\begin{figure}[t]
\centering
\includegraphics[height=8cm,width=9cm]{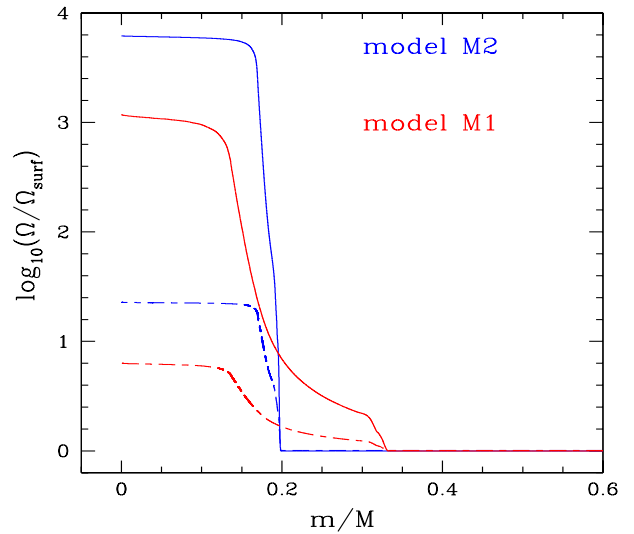}
\caption{Rotation profiles  as a function of the normalized mass.
   The surface values are  $\boldsymbol \Omega\ind{surf}/(2\pi)=0.154$, $0.042 ~\boldsymbol \mu$Hz 
   respectively. 
 Original $\Omega$ profiles  stemmed from the evolution of the stellar models
  (red solid curve for  model M1 and  blue solid curve for  model M2).
  The dot-dashed lines represent the rotation  profiles  
  that have been arbitrarily decreased  by assuming
   $\Omega\ind{surf}(\Omega/\Omega\ind{surf})^{1/ f}$
 with  $f=3.85$ for model M1 (red curve)  and $f= 2.8$ for model M2 (blue curve).
  }
\label{rot}
\end{figure}

 Model M2 is more evolved and more centrally condensed than M1.
  Accordingly,  the  inner maximum  of the Brunt-V\"ais\"al\"a frequency is 
  higher  than for the younger  model M1 \citep{2010ApJ...721L.182M} 
  and $N$ decreases more sharply. The range of frequency corresponding to the expected excited modes are indicated by horizontal lines. 
  They show that the evanescent region between the inner g resonant cavity and the upper p resonant  cavity (\emph{i.e.}, between the Brunt-V\"ais\"al\"a  and Lamb frequencies) is quite narrow for these modes.

\subsection{Rotation profiles}
 
The original rotation profiles obtained for  models M1 and M2 from evolution are displayed in Fig.~\ref{rot}. 
The central regions of the more evolved model M2 are rotating much faster than those of M1 
(amounting to 181.5 and 260 $\mu$Hz respectively). 
With such high rotation rates, we  checked  first that  the centrifugal acceleration  is negligible
compared to local gravity  in our models  (that is, with a relative magnitude of  $10^{-4}$ at most). 
 We can therefore  still safely assume  a spherically   symmetric equilibrium state.  
However,  the linear approximation is no longer
valid  due to Coriolis effects when  $2\Omega/\omega $ is close to unity. This happens in the central
regions of the models for the modes we consider. 
As we study here  slowly rotating red giants, we assumed that some slowing down 
process has and/or is occurring  in real stars that are not yet included in our models. 
We therefore artificially decrease the rotation rates stemmed 
from our evolutionary models so that the first-order approximation for the rotational 
splittings is valid. The shape of the rotation profile due to slowing down is unknown, we therefore 
keep it unchanged and investigate whether this can be
confirmed or not by confrontation with observations. 

We therefore decrease the central rotation rates by rescaling 
the rotation profile stemmed from our evolutionary models, decreasing the central values 
but keeping the  same envelope rotation rate.  
The decrease is set so that the value of the computed rotational splittings
 agree with  observed ones. For instance, for model M1, 
 the  initial  central rotation,  181.5 $\mu$Hz,  is decreased down to 0.967 $\mu$Hz. 
 The rotation profiles we use to compute the linear
rotational splittings in the following sections are then shown in Fig.~\ref{rot}.

\section{Classification of mixed modes}
\label{class} 

The  rotational splittings depend not only on the rotation profile but also on the 
eigenfunctions -- that is on the physical nature -- of the excited modes.  
Stellar models of red giant stars exhibit an outer convective region that is able to  
efficiently drive modes  stochastically \citep{2012A&A...543A.120S}. 
Thus, we consider the frequency ranges for our models that span an interval 
of a few radial orders below and above $n\ind{max}$ (the radial order 
corresponding to the frequency at maximum power).   We estimate $n\ind{max}$ 
as $n\ind{max} = \nu\ind{max}/\Delta \nu$, where the frequency at maximum power 
spectrum $\nu\ind{max}$ and the mean large separation $\Delta \nu$ are given by 
the usual scaling relations \citep[e.g., ][]{1995A&A...293...87K}
\begin{eqnarray}
\nu\ind{max}  &= & \nu_{{\rm max},\odot}  ~\left(\frac{M}{M_\odot} \right)~\left(\frac{R}{R_\odot}\right)^{-2}\left(\frac{T\ind{eff}}{T_{{\rm eff},\odot}} \right)^{-1/2} \, , \\
\Delta \nu &=&  \Delta \nu_{\odot} ~ \left(\frac{M}{M_\odot}\right)^{1/2}~\left(\frac{R}{R_\odot} \right)^{-3/2}\, , 
\end{eqnarray}
with $T_{{\rm eff},\odot}=5777$ K 
and the reference asymptotic values $\nu_{{\rm max},\odot} = 3106 \,\mu$Hz, 
 and $\Delta \nu_{\odot}=138.8 \,\mu$Hz defined and calibrated by  \cite{mesure}.
Indeed, it has been conjectured by \cite{1991ApJ...368..599B}, then shown observationally by \cite{2003PASA...20..203B}, that  $\nu\ind{max}$ 
predicts well the location of the excited frequency range of stochastically excited 
modes. A physical justification has been recently proposed by \cite{2011A&A...530A.142B}. 
The adiabatic oscillation frequencies  and eigenfunctions for $\ell=1$ modes are computed with the ADIPLS code \citep{2008Ap&SS.316..113C}. 

\begin{figure}[t]
\centering
\includegraphics[height=9cm,width=9.5cm]{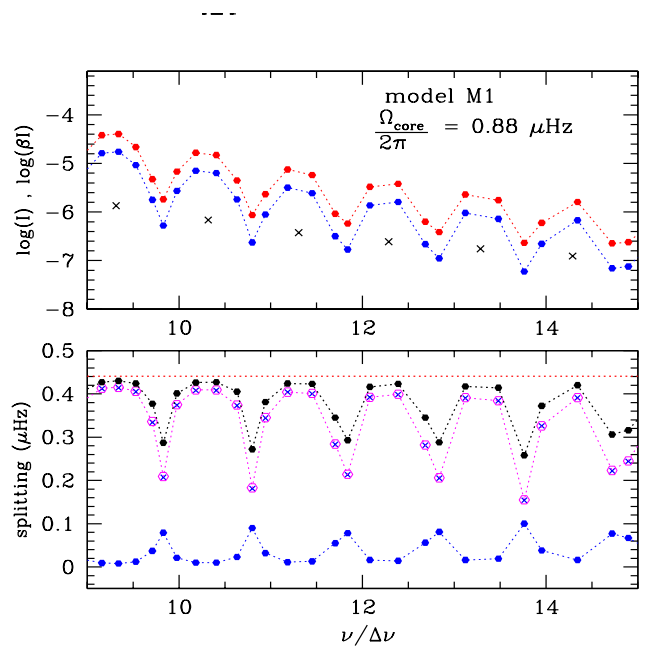}
\caption{
{\it Top:} 
Integrated kernel  ($ \beta ~  I$) with $\beta$ 
and $I$ respectively given by Eq.~\ref{eq:beta} 
and Eq.~\ref{eq:iner1}
  (blue dotted line, dots correspond to modes) 
 and mode inertia  $I$  (red dotted line, dots correspond to modes)  in decimal logarithm
 for model M1. 
 Black crosses represent the inertia of radial modes.
  {\it Bottom:} Rotational splittings (Eq.\ref{eq:split}),  in $\mu$Hz,  for $\ell=1$ modes 
 as a function of the normalized frequency  $\nu/\Delta \nu$ 
 for M1 (black dots connected by a dotted line).
Dotted blue and  magenta curves  and blue crosses
 represent various contributions and approximations to the
rotational splittings
(see text Sect.\ref{seismicconstraints}).
 The value  $(1/2)\Omegacore /2\pi $ is represented by the horizontal red dotted line.
}
\label{rot2a}
\end{figure}

As expected from previous works  \citep[e.g.,][]{1975PASJ...27..237O,1977AcA....27...95D,
 2001MNRAS.328..601D,2009A&A...506...57D,2010ApJ...721L.182M}, our red giant models show many mixed $\ell=1$ modes. For model M1, for instance, the frequency range of excited modes is expected to be 
$ \log_{10} (\sigma^2) \sim 2-2.8$. For modes that are expected to be excited around $\nu_max$ (given in Table~\ref{Tab2}), 
the corresponding frequencies are much smaller than both the Brunt-V\"ais\"al\"a frequency
and the Lamb frequency in the central regions of the models (see Fig.~\ref{vais1}). Therefore, modes
with frequencies close to $\nu_{max}$ are trapped in the g resonant cavity, located in the
radiative central region. Moreover, $\sigma_{\rm max}^2 \gg N^2, S_{\ell}^2$ in most of
the envelope so that the modes are also trapped in the p resonant cavity in the convective envelope 
(not shown). These cavities boundaries are delimited by turning points, i.e. the radii satisfying $\sigma^2 = N^2(r)$
(g cavity bounded by x$_1$ and x$_2$), or $\sigma^2 = S_{\ell}^2(r)$ (p cavity
bounded by x$_3$ and x$_4$). The location of these turning points for a list of selected
modes are given in Sect. 4-1, Table~\ref{Tab3}.

 In order to interpret correctly the observed rotational splittings, we need to 
 define a measure of the g nature of the modes with respect to their p part  and classify
  them accordingly. 
 Convenient quantities for that purpose are the kinetic energy  \citep{Unno89} or 
 the  mode inertia \citep{2001MNRAS.328..601D}, defined as
\begin{equation}
\label{inertia}
{\cal I}= {4\pi R^3}~ {\int_0^1 ~\Bigl(\xir^2+\Lambda \xih^2\Bigr)~\rho~x^2~ {\rm d}x} \, , 
\end{equation}
with  $\Lambda = \ell(\ell+1)$, and $x=r/R$ is the normalized radius. 
The oscillation quantities entering the above definition are the fluid vertical and horizontal 
displacement
 eigenfunctions $\xir, \xih$ respectively. 
For later convenience, we will use  the following variables
 \begin{eqnarray}
 \label{newvar1}
z_1 &=& \left(\frac{3 \rho }{\bar{\rho}}\right)^{1/2}~ x^{3/2} ~\frac{\xir}{R} \, , \\
 \label{newvar2}
z_2 &=& \sqrt{\Lambda} ~\left(\frac{3 \rho }{\bar{\rho}}\right)^{1/2} ~x^{3/2}~\frac{\xih}{R}\, . 
\end{eqnarray} 
Using Eqs.~(\ref{newvar1}) and (\ref{newvar2}) into (\ref{inertia})
 leads to (dropping the subscripts $n,\ell$) 
\begin{equation}
I  = \int_0^1 \Bigl(z_1^2+z_2^2\Bigr) ~\frac{ {\rm d}x}{x}  \, ,
\label{eq:iner1}
\end{equation}
where we have defined the dimensionless mode inertia $I$ as $I=  {\cal I}/(M R^2)$.

\begin{figure}[t]
\centering
\includegraphics[height=9cm,width=9.5cm]{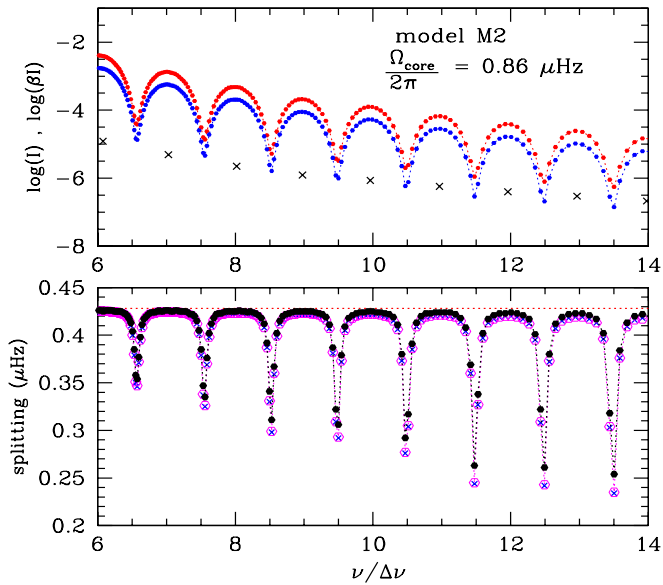}
\caption{Same as Fig.~\ref{rot2a} but  for  model M2.  }
\label{rot2b}
\end{figure}

 The variation of $\ell=1$ mode inertia with frequency is shown in Figs.~\ref{rot2a}
  and \ref{rot2b} for models M1 and M2. The number of nodes in the g cavity is given by  
 \begin{equation}
n_g = \frac{\sqrt{\Lambda}}{\pi \omega}~ \int\ind{core} \frac{N}{ x} ~dx \, , 
\label{eq:ng}
 \end{equation} 
\citep[][and references therein]{Unno89}. 

Because  $N^2$ increases with evolution (see Fig.~\ref{vais1} for M1 and M2),  the
number of nodes, hence of modes, in a given frequency range increases with the age of the star.  
This can be seen by comparing
Fig.~\ref{rot2a} for model M1 and Fig.~\ref{rot2b} for 
model M2. Mixed modes with a g-dominated character have their inertia larger than  
 modes that have their inertia shared between the g  and p cavities. 
These results are in agreement with previous works \citep[e.g.,][]{2009A&A...506...57D}. 

We now define for convenience  a measure  of the  g nature of the mode 
with the ratio of mode inertia in the g cavity 
over the total mode inertia \citep{2012ApJ...756...19D}, \emph{i.e.}, 
\begin{equation}
\zeta = \frac{\Icore}{I} = \frac{1}{I}  ~\int_0^{r\ind{core}} (z_1^2+z_2^2) ~\frac{{\rm d}x}{x} \sim 
\frac{1}{I} ~\int_0^{r\ind{core}} ~z_2^2 ~ \frac{{\rm d}x}{x} \, .
\label{eq:zeta}
 \end{equation} 
For a  mode  with a  p-dominated nature, the inertia is concentrated in the p-cavity 
and is  small due to the low density in the envelope ($\zeta \sim 0$). 
Its frequency is mainly determined by the p-cavity and is therefore half a large separation away from the frequencies of consecutive radial modes. 

For a mixed  mode where the g character dominates ($\zeta \sim 1$ ), 
the inertia is large due to the high density of the central regions where the mode has its maximum
amplitude. The frequency of this $\ell=1$ mixed mode differs significantly 
from that of a $\ell=1$ mode that would be a pure p mode. 
Its frequency is then closer to that of the closest radial mode. 
We refer later to such a mode as a {\it g-m  mode} as in \cite{2012arXiv1209.3336M}. 

When the  contributions to mode inertia  from the core and the envelope 
are nearly equal,  the mode frequencies are  less affected by the core 
and therefore remain closer to the frequencies of pure $\ell=1$ p modes. 
We will refer to such mixed modes as {\it p-m modes} \citep{2012arXiv1209.3336M}. 
These modes take intermediate values of $\zeta$. 

\section{Linear rotational splittings}\label{splitting}

For slow rotation, a first-order perturbation theory provides the following 
expression for the rotational splittings \citep{1951ApJ...114..373L,1991sia..book..401C}
\begin{equation}\label{eq:split}
\delta \nu =  \int_0^1 ~K(x) ~\frac{\Omega(x)}{2\pi}~ {\rm d}x \, , 
\end{equation}
where  $\Omega$ is the angular rotation velocity and the rotational kernel $K$ takes the form
\begin{equation}
\label{eq:K}
 K =\frac{1}{I}\Bigl(z_1^2+ z_2^2- \frac{2}{\sqrt{\Lambda}} z_1~z_2- \frac{1}{\Lambda}~z_2^2\Bigr) ~ \frac{1}{x}
\end{equation}
normalized to the mode inertia $I$. For later purpose, we also define 
 \begin{equation}
 \beta =\int_0^1 K(x)~ {\rm d}x \, .
\label{eq:beta}
\end{equation}

\begin{figure}[t]
\centering
\includegraphics[height=9cm,width=9.5cm]{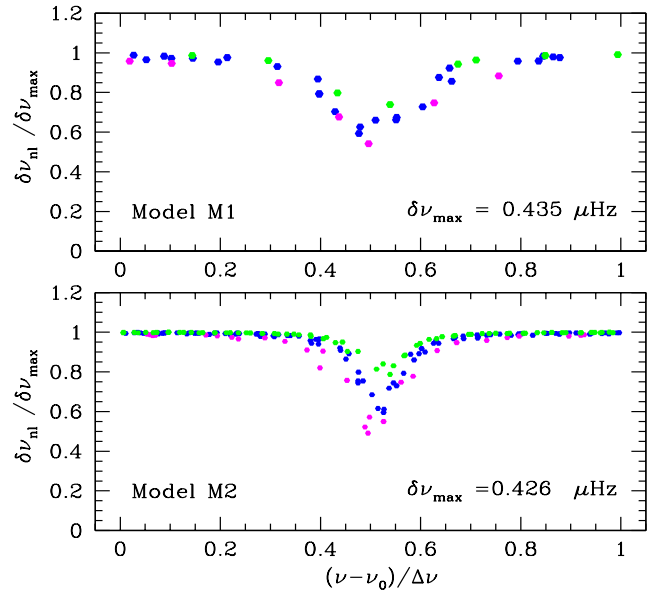}
\caption{ Folded rotational splittings according to $(\nu-\nu_0)/\Delta \nu$ 
for  {\it Top:} model M1 (blue $9<\nu/\Delta \nu <15$; magenta $\nu/\Delta \nu> 15$;
 green $\nu/\Delta \nu< 9 $) and
{\it Bottom:} for model M2 (blue $8<\nu/\Delta \nu <14$; magenta $\nu/\Delta \nu> 14$; 
green $\nu/\Delta \nu< 8 $). 
 The splittings are normalized to their maximum values, which are written in the panels.  
 }
\label{fold1a}
\end{figure}

We compute the rotational splittings  according to Eq.~(\ref{eq:split}) for  
$\ell=1$ modes and will refer to them as numerical rotational splittings later on. 
The variations of the rotational splittings with  the
normalized frequency $\nu/\Delta \nu$ 
are shown for model M1 an M2  in Figs.~\ref{rot2a} and \ref{rot2b}. 
Also displayed are  the variations  of both the
inertia $I$ and the  product
 $\beta ~ I$ (with $\beta$ 
and $I$ respectively given by Eq.~\ref{eq:beta} 
and Eq.~\ref{eq:iner1}).  
They show that the  oscillatory variations of the splittings with 
frequency are dominated by  the nature of the modes as
 they closely follow those of mode inertia and  of the integrated kernel
  $\boldsymbol \beta ~\mathbf I$ (see Sect.~\ref{pattern1}).
  
   It may be useful to the reader to note that models that are located at the same location  
in the HR diagram (models with the  same mass but one  with an overshoot of 0.1 pressure scale 
height and no rotational induced mixing and one without overshoot but with rotational induced mixing) 
 have similar structures. As a consequence, the  rotational 
 splitting of any given mode in the range of interest here 
  is the same for both models when computed  with the
eigenfunctions of the respective models and the same rotation profile.

\subsection{Periodic patterns for rotational splittings}
\label{pattern}

It is clear from Fig. \ref{rot2a} and \ref{rot2b} that the variation of the mode inertia and rotational 
splitting with frequency shows a quasi-periodic pattern, with a period close to the large separation.
Each pattern around a radial mode carries roughly the same information about the rotation, at least
 in the range of observed modes for red giants as found by \cite{2012arXiv1209.3336M}. Note, however, 
 that at higher frequencies, mixed modes become closer to pure p modes and probe 
 the rotation of the surface layers,
 while at lower frequency, the nature of g-m modes is much 
closer to that of pure g modes and probe better the inner rotation.
    
The existence of repeated patterns leads us to gather  information by folding the power spectrum according to the large separation $\Delta \nu$. The rotational splittings are plotted, in Fig.~\ref{fold1a}, as a function of $(\nu-\nu_0)/\Delta \nu$ where $\nu_0$ is the closest radial mode to the $\ell=1$ mode of frequency $\nu$.
 The folding $(\nu-\nu_0)/\Delta \nu$ corresponds to the asymptotic folding 
$\nu/\Delta \nu -(n_p+\varepsilon)$ as suggested by \cite{2012A&A...540A.143M}, 
where $n_p$ is the radial order of the mode and $\varepsilon$ represent some departure from the asymptotic description \citep{1980ApJS...43..469T,2012arXiv1209.3336M}. This folding is shown for 
our two models in Fig.~\ref{fold1a}. We see that all  patterns do nearly superpose for each model. 
Here $\varepsilon \sim 1.3-1.4$ for model M1 and $\varepsilon \sim 0$ for model M2. 
A clear correlation exists between the magnitude of the  splitting and 
the  g nature of the mixed mode. The rotation splittings are large for g-m modes 
that lie on the left and right sides of a minimum occupied by the 
p-m modes that have the smallest  rotation splittings. This is in agreement 
with recent observations \citep{2012A&A...540A.143M}. 
 In the low-frequency domain, the modes are,
  for most of them, g-m modes. At higher frequency, 
  they can be either g- or  p-m modes 
 (Fig.~\ref{fold1a}). This is clearly seen when 
 one looks at restricted domains of frequency (corresponding to different color 
 codes in Fig.\ref{fold1a}). 

We also note that the ratio of  the minimum splittings $\delta \nu\ind{min}$ to the maximum ones $\delta \nu\ind{max}$ ($\sim 0.5$ for for model M1 and M2) is only slightly sensitive to rotation. 
Actually, it depends  only on the ratio of the envelope to core rotation which is small.  This is explained in Sect.\ref{interpretation}.

\begin{table}
\caption{Model M1 frequencies of the selected pattern: 
the frequency  $\nu$ in $\mu$Hz, 
 $\boldsymbol\sigma$ is the normalized frequency $2\pi \nu (GM/R^3)^{-1/2}$, 
 $I$  is the mode inertia arbitrarily
 normalized to  mode inertia of the radial mode $n=8$; $\zeta=\Icore/I$}
\centering
\begin{tabular}{c c c c c c c}
\hline
\hline
             &        &      & $\ell=0$                    &         &\\
\hline
    $n_p$  & $\nu$ & $\sigma^2$ & $\nu/\Delta \nu$  &$\zeta$ &  $I  (10^{-6})$\\
\hline
  10    &   261.33 &   233.33 & 11.31  & $6.09~10^{-4}$  &0.375\\
\hline
\hline
             &        &     &  $\ell=1$                       &         &\\
\hline
freq. number   & $\nu$ & $\sigma^2$ &  $\nu/\Delta \nu$ & $\zeta$ &$I ~(10^{-6})$\\
\hline
$ \#1   $ & 249.62& 212.88 &    10.80&  0.422&  0.867\\
$ \#2   $ & 252.93& 218.58 &    10.94 & 0.799&  2.333\\
$ \#3   $ & 258.52& 228.34 &    11.18&  0.937&  7.512\\
$ \#4   $ & 264.70& 239.37 &    11.45&  0.928&   5.759 \\
$ \#5   $ & 270.52& 250.02 &    11.70 &  0.657&  0.922\\
$ \#6   $ & 273.65& 255.84 &    11.84&  0.498&  0.578\\
\hline
\hline
\end{tabular}
\label{Tab2}
\end{table}

\begin{table}
\caption{Normalized radius of the turning points 
for selected modes  for  model M1 frequencies: 
the g-cavity is delimitated by $[x_1, x_2]$ and the p
cavities by $[x_3, x_4]$. In the other regions the modes are evanescent. We find
  $x_1=1.3~10^{-4}$ and $x_4=0.9999$ for all  listed modes }
\centering
\begin{tabular}{c c c c c c c}
\hline
\hline
     &    $x_2$ & $x_3$ \\
\hline
\#1    &      0.0732 &   0.119     \\
\#2    &      0.0725  &  0.118    \\
\#3    &      0.0715 &   0.116  \\
\#4    &      0.0705   & 0.113   \\
\#5    &      0.0693 &   0.111    \\
\#6    &      0.0689 &   0.110 \\
\hline
\hline
\end{tabular}
\label{Tab3}
\end{table}

\subsection{Properties of a periodic pattern}
\label{pattern1}

To interpret the behavior of the splittings, we consider a single pattern around 
a given $\ell=0$ mode with radial-order $n_p =10 $ (Table~\ref{Tab2}) 
for model M1. There are six modes in the pattern labelled  $\nu_1$ to $\nu_6$ 
from the smallest to the largest frequency.

\subsubsection{Mode inertia $I$ and rotational splittings }

\begin{figure}[t]
\centering
\includegraphics[height=9cm,width=9.5cm]{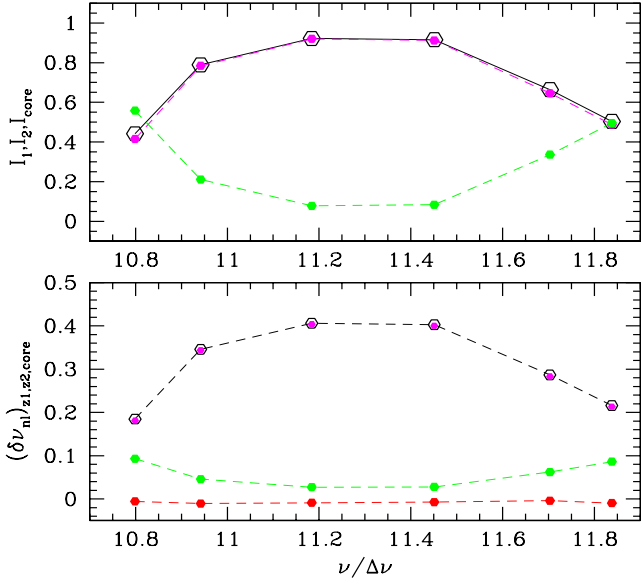}
\caption{{\it Top:} relative inertia contributions $I_1/I$ 
(green dots connected dashed line dots correspond to modes) and $I_2/I$  (Eq.\ref{eq:I1}) 
 (black),   relative  contribution $\zeta= \Icore/I$ 
 (magenta dashed line,  hexagons correspond to modes).
 {\it Bottom:}  contribution to the rotational splittings for model M1 
due to  $z_1^2$  only 
(green dashed line dots correspond to modes),   contribution due to  $z_2^2 $ only
 (black  dashed lines and hexagons),   contribution  due to  $ z_1 z_2$  only 
 (red-dashed line, dots correspond to modes); 
 contribution to $\delta \nu_{nl}$ due to the core only
 (magenta filled  hexagons) 
 }
\label{Itot}
\end{figure}

The rotational splittings  for $\nu_1$ to $\nu_6$ are displayed in Fig.\ref{Itot}. 
The relative contributions $I_1/I$ 
related to  $z_1^2$, and $I_2/I$ related to $z_2^2$, defined as  
\begin{equation}
 \frac{I_1}{I} = \frac{1}{I}  ~\int_0^{1} z_1^2 ~\frac{{\rm d}x}{x} ~~~;~~~
 \frac{I_2}{I} = \frac{1}{I}  ~\int_0^{1} z_2^2 ~\frac{{\rm d}x}{x} 
\label{eq:I1}
 \end{equation} 
are displayed in Fig.\ref{Itot}. 

The contribution  of the $z_1 z_2$ terms to the  rotational splittings (Eq.\ref{eq:split}) 
is negligible in front of $z_1^2$ in the envelope and in front of $z_2^2$ in the core. Thus, 
the following conclusions apply for both the mode inertia and the rotational splittings. 

The $I_1$ contribution coincides to a good approximation with the envelope contribution
$I\ind{env}$, while the $I_2$  contribution coincides  to a good approximation with the 
  core contribution. The core contribution to inertia, $\Icore$, 
  and  the rotational kernel,  $K_{\Omega,core}=\int_{core} \Omega(x) ~K(x) ~{\rm d}x$, are 
computed between the  
lower and upper turning points of the g cavity for each mode. 
The  dense core contributes heavily 
  to the total mode inertia $I$  and rotational splittings.  
  The values of the ratio $\zeta= \Icore/I$  
   are given in Table~\ref{Tab2}.  

\begin{itemize}
\item The modes with frequencies $\nu_3$ and $\nu_4$ are g-m  modes.
 For these modes, the contribution of the horizontal
displacement  $z_2$ (essentially in the core) dominates their inertia ($I_2/I \sim \Icore/I \approx 1$) 
while the contribution due to the radial displacement $z_1$ (largest in the envelope), $I_1/I$,  
is negligible.  

The envelope above the core where the mode is either evanescent or of acoustic type 
contributes for almost nothing to the total mode inertia $I\ind{env}/I= 1-\zeta \ll \zeta$ 
for such modes.
 The contribution of the envelope to $\delta \nu$ is negligible  because the rotation 
 is quite small  there.

\medskip
\item  The other modes are p-m modes. 
 The envelope above the core where the mode is either evanescent or
   of acoustic type contributes  for half 
   for these modes ($I_1\sim I_2 \sim I/2$).  
They  share their inertia almost equally
 in the core and in the envelope  with an almost equal contribution of the horizontal 
 and radial displacements.

\medskip
     
\end{itemize}

\subsubsection{Integrated kernels}

\begin{figure}[t]
\centering
\includegraphics[height=9cm,width=9.5cm]{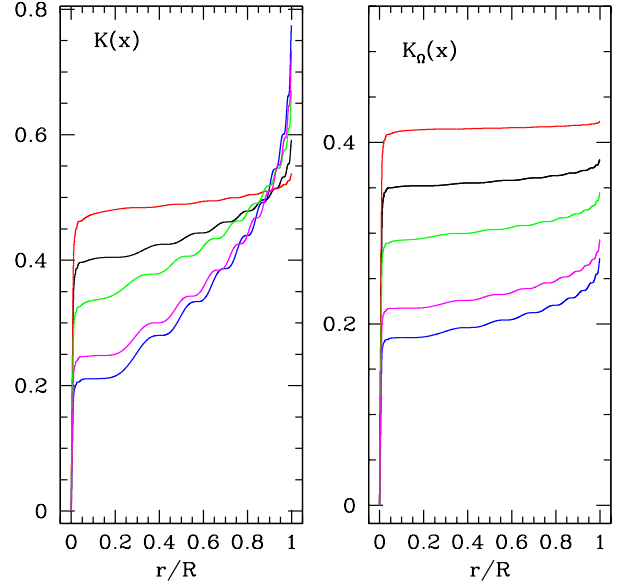}
\caption{ Integrated kernels $  \int_0^x~K(x')~ {\rm d}x'$ (left panel) and 
$\int_0^x~\Omega(x)~K(x')~{\rm d}x'$ (right panel)
(with $ K(x)$ given by Eq.~\ref{eq:K})  as a function of $x=r/R$ 
 for some modes of the  selected pattern  for model M1. 
The lines correspond to mode   \# 1 (blue); \# 2
(black); \# 4(red);    \# 5 (green)  and \# 6 (magenta) of Tab.\ref{Tab2} respectively.}
\label{Krotz} 
\end{figure}

The modes are trapped in the same g cavity, thus they  probe the same inner 
part of the rotational profile of the model. This can  be seen with the behavior of the 
integrated rotational kernels  that are  shown in Fig.~\ref{Krotz} (left panel).  
For all modes of the pattern, the integrated rotational kernels increase rapidly from the center. 

\begin{itemize}

\item  For the g-m modes, the integrated kernels saturate quite rapidly to the maximum value, indicating that the envelope does not significantly  contribute. This is the case for $\nu_4$
in our  studied pattern.  Note that the saturated value is obtained long before the  upper turning
 point of the g cavity is reached. This is due to the fact that the
 amplitudes of the displacement become rapidly small in the upper part of the g-cavity.

\medskip

\item  The p-m modes such as $\nu_1$ and $\nu_6$ have their  integrated kernels 
that saturate less rapidly and to a much smaller value than the g-m modes.
The corresponding  integrated contributions to the splittings  are shown in 
Fig.~\ref{Itot} and \ref{Krotz} (right panel). As they include the rotation profile, 
they increase faster than the integrated  rotational kernel. 
  
\end{itemize}

\section{Seismic constraint on  rotation}
\label{seismicconstraints}

 To have more physical insight into the information carried by rotational splittings it is worthwhile 
 to emphasize their dependences. To this end, we decompose the rotational 
 splittings into  two contributions.  
 Given that the inner (radiative) and outer (convective) cavities have well distinct properties,  we
 write:
 \begin{equation}
 \delta \nu =\beta\ind{core} \left<\frac{\Omega}{2\pi}\right>\ind{core} 
 + \beta\ind{env} \left<\frac{\Omega}{2\pi}\right>\ind{env} \, , 
\label{eq:split1}
 \end{equation}
where \begin{eqnarray}
\label{betacore}
\beta\ind{core} &\equiv &\int_0^{x\ind{core}} ~ K(x) ~~ {\rm d}x  \, ,  \\
 \beta\ind{env} &\equiv & \int_{x\ind{core}}^1  ~K(x) ~~ {\rm d}x = \beta-\beta\ind{core}  \, , 
\end{eqnarray}
and
\begin{eqnarray}
\Omegacore &\equiv & \frac{\int_0^{x\ind{core}} \Omega(x^\prime) ~K(x^\prime) ~~ 
 {\rm d}x^\prime}{\int_0^{x\ind{core}} ~K(x^\prime) ~~ {\rm d}x^\prime}  \, , 
\label{omcore}
\end{eqnarray}
\begin{eqnarray}
\langle\Omega\rangle\ind{env} &\equiv &
 \frac{\int_{x\ind{core}}^1 \Omega(x^\prime) ~K(x^\prime) ~~ {\rm d}x^\prime}{\int_{x_{core}}^1 ~ K(x^\prime) ~~ {\rm d}x^\prime}  \, ,  
\label{omenv}
\end{eqnarray}
where $K(x)$ is defined in Eq.~(\ref{eq:K}) and
 $\Omegacore$ is the angular rotation velocity averaged over the central layers 
enclosed within a radius $x\ind{core}=r\ind{core}/R$ corresponding to the g resonant cavity. 
The radius $x\ind{core}$  is taken as the radius of the upper turning point of the g cavity  
$x_2$ (see Table~\ref{Tab3} for one pattern). 
$\langle\Omega\rangle\ind{env}$ is defined as the angular rotation velocity averaged over the layers above
 $x\ind{core}$.

 For the modes we consider in this paper, the evanescent regions above the g resonant cavity 
 and above the p cavity are narrow. 
 Then, the envelope essentially corresponds to the p resonant cavity.  
 The boundary radii are then taken as $x_3$ and $x_4$ (see Table~\ref{Tab3}).
Strictly speaking, the above quantities  depend on the mode (actually on its frequency) 
because they are defined in  the propagative cavities characteristics of each mode. 
However, because of the sharp decrease of the Brunt-V\"ais\"al\"a frequency at the edge of the 
H shell burning region, the upper turning point of the g cavity is roughly the same for all 
modes (see Table~\ref{Tab3}). The modes have  a significant amplitude in the g cavity in a 
region that is quite narrower than the extent of the g cavity  and that  is
independent of the mode (in the frequency regime we consider). This can be seen in
 Fig.~\ref{zb}  where the horizontal and radial displacement eigenfunctions  
 are shown for two typical  modes of the  studied pattern with frequencies listed in Table~\ref{Tab2}. 
We can then consider that $x\ind{core}$  and the quantity 
 $\Omegacore$ are nearly the same for all considered modes.

The radius of the lower turning-point, $x_3$, for the p-cavity of p-m $\ell=1$ modes 
 decreases  slightly with increasing frequency of the mode while   the 
  upper turning point remains approximately the same (see Table.~\ref{Tab3}). 
Thus, mode inertia  remains dependent on the mode. On the other hand, 
 the mean envelope rotation is nearly that given by the uniform 
 rotation of the convective zones and remains  the same for  all modes. 
\begin{equation}
\langle\Omega \rangle\ind{env} \approx \Omega\ind{CZ}  \, .
\label{omcoreg}
\end{equation}

\begin{figure}[t]
\centering
\includegraphics[height=9cm,width=9.5cm]{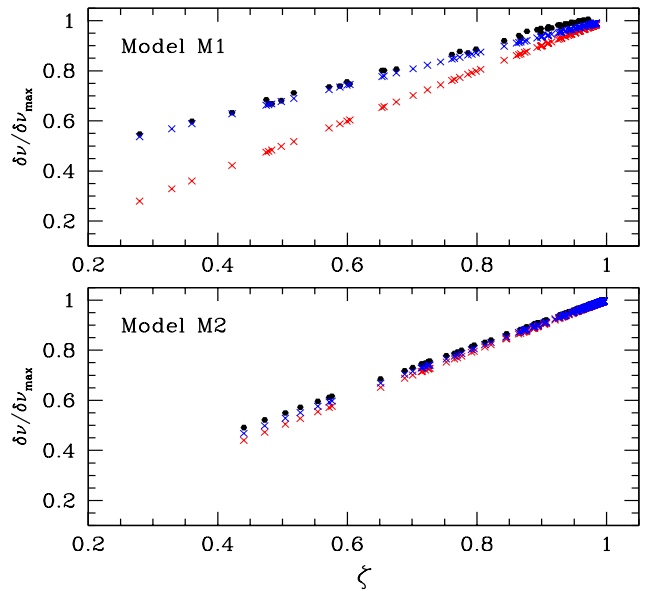}
\caption{The  rotational splitting normalized to its maximum value  as a
function of $\zeta$ (black dots) for model M1 (top) and model M2 (bottom).
 The red crosses represent  the
approximation $\delta \nu/\delta \nu\ind{max}=  \zeta$ as a function of
$\zeta$. The blue crosses represent the
approximation $\delta \nu/\delta \nu\ind{max}=  \zeta (1-2{\cal R})+2{\cal R}$ 
as a function of
$\zeta$ with  $R=0.1785$ and using numerical values of $\zeta$. }
\label{fig:omtaug} 
\end{figure}

In Fig.~\ref{rot2a}, the rotational splittings obtained with 
two approximations for $\Omegacore$ are compared  with the exact calculation 
Eq.~(\ref{eq:split}).
The core contribution to the rotational splittings $\Omegacore \beta\ind{core}$ 
 (Eqs.~(\ref{eq:split1}) and (\ref{omcore})) 
 is displayed with the magenta dots connected with the magenta dotted line. The blue crosses dotted line represents the core contribution to the 
rotational splittings  (Eq.~(\ref{eq:split1})) with $\Omegacore$ computed with only the horizontal eigenfunction
$z_2$. The two variations (magenta dots and blue crosses) cannot be distinguished. 
The blue dotted line represents the  contribution from the envelope to the 
rotational splittings $\langle\Omega\rangle\ind{env} ~(\beta-\beta\ind{core})$. 
This contribution is  much smaller for model M2 than for model M1.

For the   g-m modes (i.e. for $\zeta \sim 1$), $\beta\ind{core} \approx 1/2$.
The maximum rotational splitting is then given by 
\begin{equation}
\delta \nu_{\rm max}= \frac{1}{2}~ \left<\frac{\Omega}{2\pi}\right>\ind{core} \, .
 \end{equation}
The maximum values of the computed rotational splittings, $\delta  \nu\ind{max}$, 
 reach  $0.44$  and  $0.430~\mu$Hz    for  M1 and M2
models respectively. These values give rise to a mean core rotation 
$  \left<\Omega/2\pi\right>\ind{core} = 0.88~\mu$Hz and  $0.86~\mu$Hz, which are close to  but 
smaller than the central rotation for models M1 and  M2 respectively 
(resp.  0.97 and 0.95\,$\mu$Hz).  This is explained by the fact that the rotation
decreases sharply in the central regions of our models
and that the major contribution to the rotational splitting 
is slightly shifted off the center as are the maximum amplitudes of the horizontal 
displacement eigenfunctions 
(Fig.~\ref{zb}) and the rotational kernels.
  
Moreover, for $\ell=1$ modes, one
obtains $\beta\ind{core} \simeq \zeta/2$ , $\beta = 1 - \zeta/2$, and
$\beta\ind{env}= 1-\zeta$, to a very good approximation  (Eq.~(\ref{eq:betacore1}) in Appendix \ref{beta}).
   Then
 the linear rotational splitting Eq.~(\ref{eq:split1}) is easily rewritten as 
 \begin{equation}
\frac{\delta \nu}{\delta \nu\ind{max}} =    \zeta   ~\left(1 - 2 {\cal R}\right) + 2 {\cal R} \, , 
\label{eq:ratio}
 \end{equation}
where we have defined the ratio 
\begin{equation}
{\cal R} \equiv \frac{ \langle\Omega \rangle\ind{env}}{\Omegacore} \, .
 \end{equation}
  The splittings linearly  increase with $\zeta$. This linear dependence 
   is verified  using the numerical frequencies  computed for 
   model M1 and M2 and is  shown in Fig.~\ref{fig:omtaug}.

We stress that  for model M2 the ratio  ${\cal R}$ 
imposed by our choice of rotation profile  is quite small and the approximation
\begin{equation} 
\label{delnu2}
 \delta \nu \approx \frac{1}{2} \left<\frac{\Omega}{2\pi}\right>\ind{core} ~\zeta
 \end{equation}
is in agreement with the numerical splitting values.

We find that the rotational splittings of p-m modes as well as g-m modes are dominated by the  central layers for model M2. For the most g-dominated modes, $\zeta \sim 1$ and the rotational splittings in that case directly give half the  mean core rotation. 

 This is not true for less-evolved  models such as model M1 for which the approximation Eq.~(\ref{delnu2}) is not sufficient (Fig.~\ref{fig:omtaug}). The contribution from the envelope layers is not negligible
 for the rotation profiles we have assumed and the surface rotation must be taken into account as in Eq.~(\ref{eq:ratio}). This is explained by the fact that the p cavity extends quite deep downward inside  the model  where  the rotation already  shows  a sharp  gradient toward the center.
 
\begin{figure}[t]
\centering
\includegraphics[height=9cm,width=9.5cm]{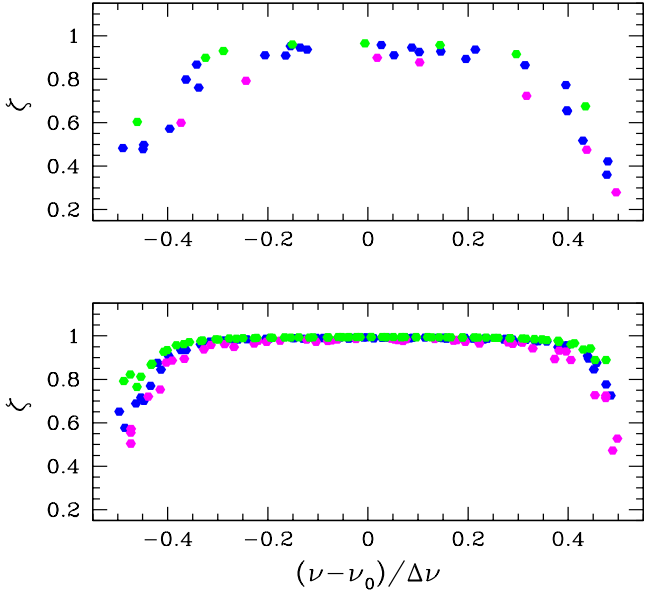}
\caption{$\zeta= \Icore/I$  as a function $(\nu-\nu_0)/\Delta \nu$ for all $l=1$
 modes for model M1 (top panel) and for model M2 (bottom panel).
$\nu_0$ is the  frequency of the closest radial mode for  the l=1 mode with frequency $\nu$.
For model M1, blue dots correspond to  modes with 
 $9<\nu/\Delta \nu <15$; magenta dots to   modes with $\nu/\Delta \nu> 15$ and
 green dots to modes with $\nu/\Delta \nu< 9 $).
For model M2, blue dots correspond to  modes with 
 $8<\nu/\Delta \nu <12$; magenta dots to   modes with $\nu/\Delta \nu> 12$ and
 green dots to modes with $\nu/\Delta \nu<8  $).
 }
\label{replizeta}
\end{figure}

The interest of Eq.~(\ref{eq:ratio}) is that $\zeta$ as well as  $\delta \nu$ and  $\delta \nu_{max}$ 
 can be obtained from observations. \cite{2012A&A...540A.143M,2012arXiv1209.3336M} have shown
  that it is possible to identify the $m=0$ modes and their
nature (p- or g-dominated mixed). We show here that it is also  possible to attribute
them a  $\zeta$  value from the observations. 
We use as proxy for the measure of the g nature of the mixed mode
 the distance of the $\ell=1$ mode frequency from that of 
 its companion radial mode: $\nu-\nu_0$.
 
 Figure~\ref{replizeta} shows the $\zeta$ values 
 for $\ell=1$ modes as a function of $\nu-\nu_0$ for models M1 and M2
  where $\nu_0$ is the 
 frequency of the radial mode closest to $\nu$. 
The $\zeta$ values  all follow the same curve.
 Due to this monotonic dependence, 
 it is  possible to estimate the value of $\zeta$ from  observations (\emph{i.e.}, from the knowledge of
 $\nu-\nu_0$). The dispersion of the 
  $\zeta$ values about the mean curve in Fig.~\ref{replizeta}
   introduces uncertainties on the
 induced $\zeta$ value. This nevertheless offers a mean to derive not only 
 $\delta  \nu_{\rm max}$ -- hence the mean core rotation -- but also the  
 ratio ${\cal R}$, hence a measure of the mean
 rotation gradient from observations.
 
\section{Physical interpretation of the observed splittings} 
\label{interpretation}

In this section, we show that rotational splittings can be expressed as a function of observables. Indeed, this will further permit us to derive physical properties related to rotation by using observations only (see Sect.~\ref{Obs}). 

  We interpret physically the two components of the rotational splitting $\Omegacore$ and 
  $\langle\Omega\rangle\ind{env}$ as the rotation angular velocities averaged over the time spent 
  in the g and p cavities, respectively.  
  We checked numerically with our stellar models and numerical frequencies 
  and eigenfunctions that $\Omegacore$ and 
   $\langle\Omega\rangle\ind{env}$ are well approximated by 
     mean rotations defined as
 \begin{equation}
\bar{\Omega} = \frac{1}{\tau} \int ~\Omega (x) ~{\rm d} \tau  .
\label{eq:taupg1}
 \end{equation}
   The time spent by the mode     is approximately given by 
\begin{equation}
\tau \approx  \frac{2}{\sigma} ~ \int ~k\ind{r}~{\rm d}x \, , 
 \end{equation}
\citep[][ and references therein]{Unno89} where $\sigma$ is the  normalized frequency of the mode.
 The radial wave number of the wave normalized to the stellar radius, $k\ind{r}$, is 
given, in a local asymptotic analysis, by  
\begin{equation}\label{kr2true}
 k\ind{r}^2 = \frac{1}{\sigma^2} \frac{1}{c_s^2} \Bigl(N^2-\sigma^2\Bigr)\Bigl(S_\ell^2-\sigma^2\Bigr) \, ,
 \end{equation}
\citep{1975PASJ...27..237O,Unno89} where $N^2$ and  
 $S_l^2$ are the normalized squared Brunt-V\"ais\"al\"a and Lamb frequencies,  respectively.
As will be seen below, the level of approximation is sufficient for 
our purpose.

 In the g cavity, $\sigma^2 \ll S_\ell^2$, therefore   
\begin{equation}
k_r \approx \frac{\sqrt{\Lambda}}{\sigma x} ~(N^2-\sigma^2)^{1/2}\, , 
\end{equation}
then the time spent  in the resonant g  cavity  becomes
 \begin{equation}
 \label{taug}
\tau_{g} =  2~ \frac{\sqrt{\Lambda}}{\sigma^2}~ \int\ind{core} ~ (N^2-\sigma^2)^{1/2} ~ 
\frac{{\rm d}x}{x}  \, .
 \end{equation}
 Note that this provides the number of nodes in the g cavity  as $n_g$ 
that is defined as $ 2 \pi n\ind{g} /\sigma =\tau\ind{g}$. 
 
Similarly, in the acoustic cavity in the envelope where $\sigma^2 \gg N^2, S_l^2$, one can write 
\begin{equation}
\label{taup}
\tau_p = 2 ~ \int\ind{env} \frac{k_r}{\sigma}~{\rm d}x \approx 2 \int\ind{env} \frac{{\rm d}x}{c_s} \, .
 \end{equation}
Note that the time $\tau_p$ and $\tau_g$ are defined  in units of the dynamical time $t\ind{dyn}= (GM/R^3)^{-1/2}$. 
 
The expressions for the mean core (resp. envelope) rotation become
 \begin{eqnarray}
\bar{\Omega}_{core} &=&   \frac{1}{\tau_g} \int\ind{core} ~\Omega (x) ~\frac{N}{x}{\rm d}x \, ,
\label{eq:taupg21}
\\
\bar{\Omega}\ind{env}  &=& \frac{1}{\tau_p}  \int\ind{env} ~\Omega (x)~ \frac{{\rm d}x}{c_s} \, .
\label{eq:taupg22}
 \end{eqnarray}

\begin{figure}[t]
\centering{
\includegraphics[height=9cm,width=9.5cm]{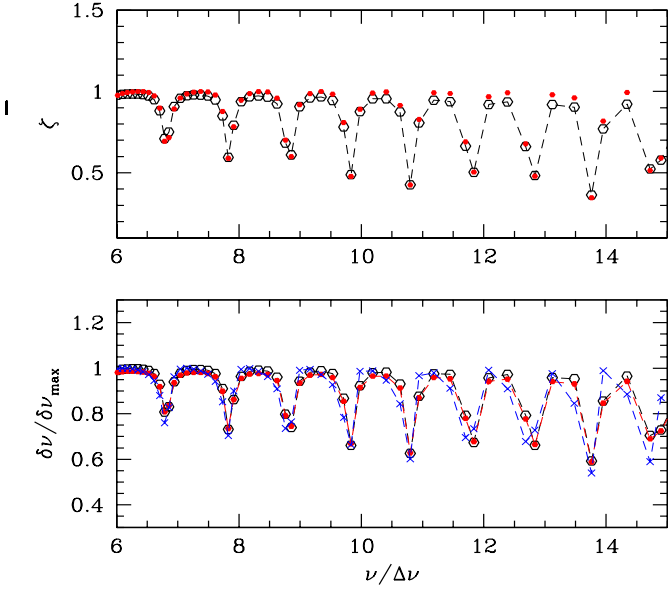}}
\caption{  
 {\it Top:} $\zeta$ as a function of $\nu/\Delta \nu$. Black open symbols:
  numerical results (Eq.\ref{eq:zeta}).
   Red dots:  $\zeta$ approximated by (Eq.\ref{eq:zeta_app}) 
   with $\chi= 2.5 ~y \cos(\pi/(\Delta \Pi \nu))$
    and $y=\nu/\Delta \nu $.  
    {\it Bottom:}  $\delta \nu /\delta \nu\ind{max}$ as  a function of 
   $ \nu /\Delta \nu$.  Black dots: numerical values (Eq.\ref{eq:split}).  Red dots: 
     approximate expression for 
    $\delta \nu / \delta \nu\ind{max}$ (Eq.\ref{eq:ratio}) 
  with ${\cal R}= 0.1785$ and $\zeta$ given by its 
numerical values.  
Blue crosses: $\delta \nu / \delta \nu\ind{max}$ given  by Eq.\ref{eq:split10}  
with ${\cal R}= 0.1785$  and $\chi$ given by  Eq.\ref{eq:chi10}.}
\label{fig:splitasM1}
\end{figure}

The  theoretical values of  $\bar{\Omega}\ind{env}$ and $\bar{\Omega}\ind{core}$, 
Eqs.~(\ref{eq:taupg21}) and (\ref{eq:taupg22}), are computed  
for both equilibrium models M1 and M2. We find  $\bar{\Omega}\ind{core}/2\pi=0.88~\mu$Hz, 
$\bar{\Omega}\ind{env}/2\pi=0.157~\mu$Hz then $ \delta \nu\ind{max,th}=0.44~\mu$Hz,
${\cal R} =0.178 $  for model M1 and $\bar{\Omega}\ind{core}/2\pi=0.856~\mu$Hz, 
$\bar{\Omega}\ind{env}/2\pi=0.042~\mu$Hz, $\delta \nu\ind{max,th}=0.428~\mu$Hz,  ${\cal R} = 0.05$
for model M2 where the subscript ${th}$ stands for theoretical.

We then computed   the rotational splittings  from  the approximate expression 
  Eq.~(\ref{eq:ratio}) using  these figures  
  with $\zeta$ computed with the associated eigenfunctions (Eq.\ref{eq:zeta}). 
  As a result, the theoretical dependence given by Eq.~(\ref{eq:ratio}) 
  perfectly matches the numerical curve $\delta \nu/\delta \nu\ind{max}$  
  and $\bar\Omega\ind{core} \sim \langle \Omega\rangle\ind{core}$,
   $\bar\Omega\ind{env} \sim \langle \Omega\rangle\ind{env}$.
 This can be seen in Fig.~\ref{fig:splitasM1} for model
M1 where one compares 
the behavior of the numerical splittings $\delta \nu/\delta \nu\ind{max}$ with
$\nu/\Delta \nu$ to the  splitting values obtained by using  Eq.~(\ref{eq:ratio}).

\medskip 

One can go to a further level of approximation by deriving an approximate expression for the mode inertia.
Appendix \ref{zeta} shows that one can derive an approximate expression for $\zeta$  as 
 \begin{equation}
 \label{eq:zeta_app}
\zeta \approx \frac{1}{1+\alpha_0 ~\chi^2} \, , 
 \end{equation}
where 
 \begin{equation}
 \label{eq:chi10}
  \chi  \approx  2~ \frac{\nu}{\Delta \nu}~ \cos\left(\frac{ \pi} { \Delta \Pi~ \nu } \right)
   = 2~ y~ \cos\left(\frac{ \pi} { \alpha_0~ y } \right)\, .
  \end{equation}
For convenience we have defined $y=\nu/\Delta \nu$  and a dimensionless constant $\alpha_0$ as
\begin{equation} 
\alpha_0 = \Delta \nu ~ \Delta \Pi \, , 
\end{equation}
with $\Delta \Pi$  is the period spacing for g-m  modes
\begin{equation} 
 \Delta \Pi = \frac{2 \pi^2}{\sqrt{\Lambda}}~  \left(\int\ind{core} ~ 
 \frac{(N^2-\sigma^2)^{1/2}}{x} ~ {\rm d}x\right)^{-1} \, .
\end{equation}
For models M1 and M2, we respectively have $\alpha_0 = 2.2~10^{-3}$ and $\alpha_0 = 6.0~10^{-4}$.

\begin{table}
\caption{Typical values for the quantity 
 $4 \alpha_0~ y\ind{n_0}^2$ (Eq.(\ref{eq:ratio6})) for stars on the RGB and in the
clump. $\Delta \nu$ and $\nu_{max}$  are given in $\mu$Hz and 
$\Delta \Pi$ in seconds. The values in parenthesis corresponds to $y_{n_0}$  
evaluated at $\nu\ind{max}\pm 3\Delta \nu$.}
\centering
\begin{tabular}{c c c c c c c}
\hline
\hline
           $\Delta \nu$  &    $\nu\ind{max}$    &    $\Delta \Pi$   &  $4 \alpha_0~ y\ind{n_0}^2$    & 	   stage       &\\
\hline
    40  & 500 & 120 &  3.0 ~(1.73, 5.23) &   bottom RGB\\
\hline
  10    & 120 &   80 & 0.46 ~(0.26, 0.82) &            mid RGB \\
\hline
  4     & 35   & 60 &  0 .074 ~(0.03, 0.16)  &  top RGB  \\
\hline
  4     & 35    & 300 &    0.43  ~(0.01, 0.09)& clump \\
\hline
\hline
\end{tabular}
\label{Tab4}
\end{table}

The behavior of the ratio $\zeta $ as a function of $\nu/\boldsymbol{\Delta} \nu$ is computed with Eq.~(\ref{eq:zeta_app}) and Eq.~(\ref{eq:chi10}) and is compared to that of the numerical one in Fig.~\ref{fig:splitasM1} (top).
 The cosine term gives rise to  the oscillation of $\zeta$  while the decrease of the minimum 
 values of $\zeta$ with frequency is driven by the ratio $\tau_p/\tau_g$.  
 $\zeta$ is
computed using Eq.~(\ref{eq:zeta_app}) and Eq.~(\ref{eq:chi10}) 
where the factor 2 is replaced by a factor 2.5
which fits better the numerical results.  This  difference is explained by the fact that we took the factor $f$ (Appendix \ref{zeta}) equal to unity while it is in  reality smaller. For model M1 the difference between the large separation  $\Delta \nu$
and the equivalent quantity computed over the p resonant cavity is of the order of 
$f=0.8$. This increases the constant in $\chi$ from 2  to 2.24, closer to the the adopted 
2.5 value. As mentionned also in  Appendix \ref{zeta}, 
the amplitude ratio Eq.~(\ref{csa}) is further increased
if one takes into account the effect of the evanescent zone between the  outer p and inner g
resonant cavities of modes. As a consequence, this leads to an increase of
 the $\chi$ term contribution.
Actually, some departure from asymptotic is expected for the p part of red giants
 modes. It is then striking that the numerical results agree that  well  
 with Eq.~(\ref{eq:zeta_app}). 

An expression for the normalized rotational splitting as a function of the observable $y\equiv 
\nu/\Delta\nu$ is then obtained by combining 
Eq.~(\ref{eq:ratio}) and  Eq.~(\ref{eq:zeta_app}) so that
 \begin{equation}
\frac{\delta \nu}{\delta \nu_{\rm max}} \approx   
\frac{1 - 2 {\cal R}} {1+ \alpha_0~\chi^2} + 2 {\cal R} \, , 
\label{eq:split10}
 \end{equation}
 with  $\chi$  given by  Eq.~(\ref{eq:chi10}).

The rotational splittings computed with Eq.~(\ref{eq:chi10})  and Eq.~(\ref{eq:split10}) are compared with
the numerical values for model M1 in Fig.~\ref{fig:splitasM1}.  
The behavior of both theoretical and numerical quantities agree quite well. 
An illustration is provided in Sect.~\ref{Obs} with the red giant star 
 KIC 5356201 \citep{2012Natur.481...55B} for which such a procedure has been used. 

The ratio $\delta \nu/\delta \nu\ind{max}$ as a function of 
$(\nu-\nu_0)/\delta \nu$ is nearly independent of rotation, in particular when ${\cal R}\ll1$. 
Maxima  of the $\zeta $ oscillation are obtained for $\cos \bigl(\pi/(\alpha_0~y)\bigr)=0$, then 
$\zeta\ind{max}= 1$. Minima are  defined for $\cos \bigl(\pi/(\alpha_0 y\ind{n_0})\bigr)=\pm 1$, 
\emph{i.e.},  $\alpha_0  y\ind{n_0} = 1/n_0$ (for any integer $n_0$) then 
 \begin{equation}
 \label{zetamin}
\zeta\ind{min} \approx \Bigl(1+4 ~ \alpha_0~ y\ind{n_0}^2 \Bigr)^{-1} \, .
 \end{equation}
 Accordingly, the contrast $\eta$ defined as the 
 ratio of minimum to maximum splittings is obtained from Eq.~(\ref{eq:split10})   using 
Eq.~(\ref{zetamin}) and Eq.~(\ref{taupstaug}) as 
 \begin{equation}
\eta \equiv \frac{\delta \nu_{\rm min}}{\delta \nu_{\rm max}} =
\frac{1 - 2 {\cal R}}{1+4 ~ \alpha_0~ y\ind{n_0}^2 } + 2 {\cal R} \, .
\label{eq:ratio6}
 \end{equation}
 As $4 \alpha_0~ y\ind{n_0}^2$ decreases with increasing integer $n_0$, the contrast
   $\delta \nu_{\rm min}/\delta \nu_{\rm max}$ increases  with
    frequency for a given rotation profile.
    This can be seen quite clearly for model M2 in
   Fig.~\ref{rot2b} for instance.
   
 Typical values of the quantity   $4 \alpha_0~ y\ind{n_0}^2$  for different 
 types of observed 
 red giants are listed
 in Tab.~\ref{Tab4}, according to the analysis done by 
 \cite{2012A&A...540A.143M}. For simplicity, $y\ind{n_0}$ is evaluated 
 at $\nu=\nu\ind{max}$. 
 The first two  lines are given for red giants  corresponding well  to 
 models M1 and M2 
 respectively discussed in the previous sections.
More generally,   this quantity 
decreases from about 4 at the bottom of the  RGB down to 0.06  
for red giants stars located on the highest part of the RGB 
 for which  stochastically oscillations  
are detected. It amounts to roughly 
 ~0.4  for the clump stars.
 Note that for the most evolved red giant stars, $4 \alpha_0~ y\ind{n_0}^2$ is quite small.
  In that case, the contrast $\eta$ varies as 
 \begin{equation}
\eta \equiv \frac{\delta \nu_{\rm min}}{\delta \nu_{\rm max}}  \sim  1-4 ~ \alpha_0~ y\ind{n_0}^2 ~ (1-2{\cal R})\, .
\label{eq:ratio7}
 \end{equation}
and the influence of the rotation on this contrast throughout ${\cal R}$  can become negligible.

\begin{figure}[t]
\centering{
\includegraphics[height=6cm,width=8.cm]{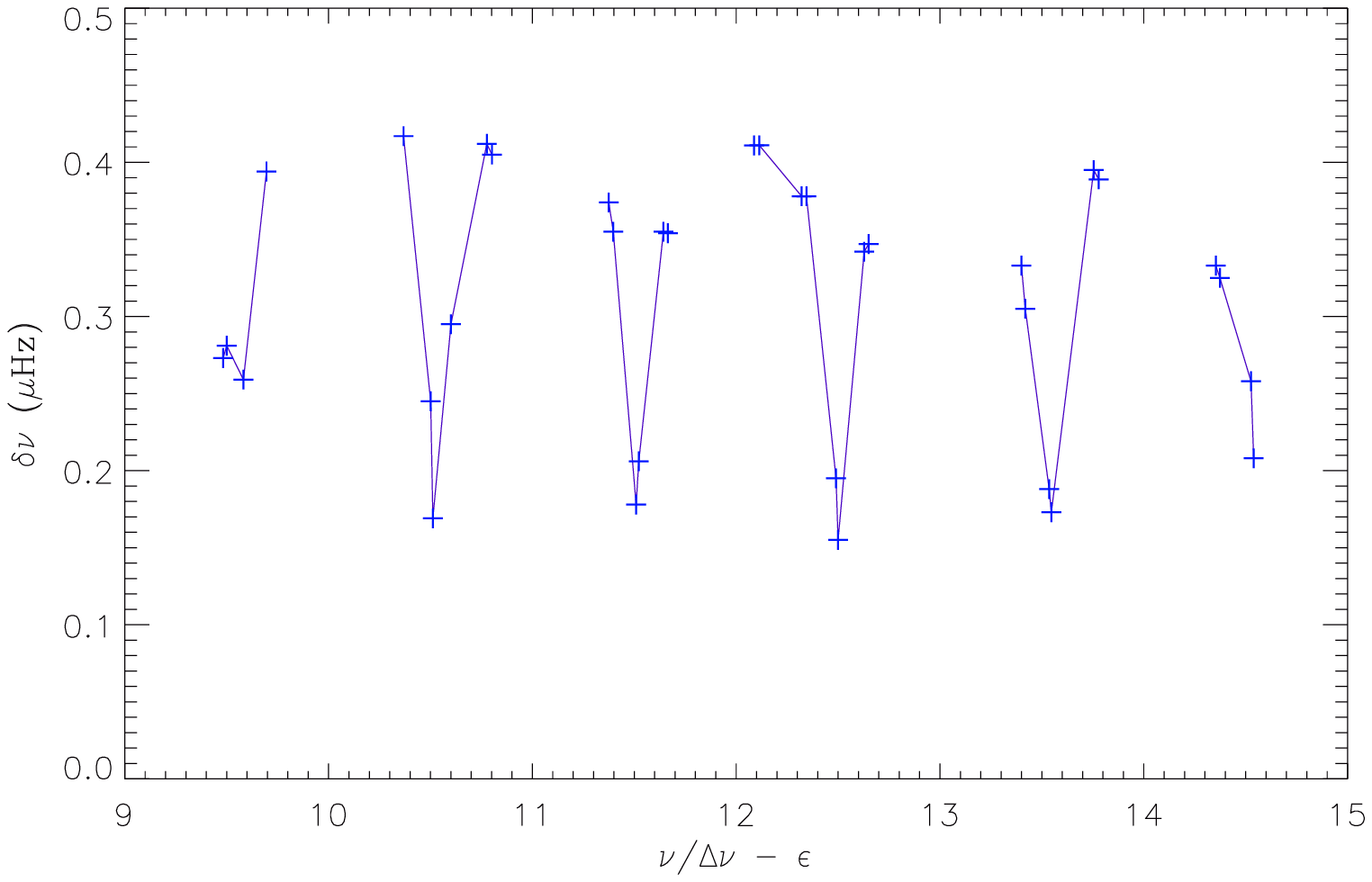}\\
\includegraphics[height=6cm,width=8.4cm]{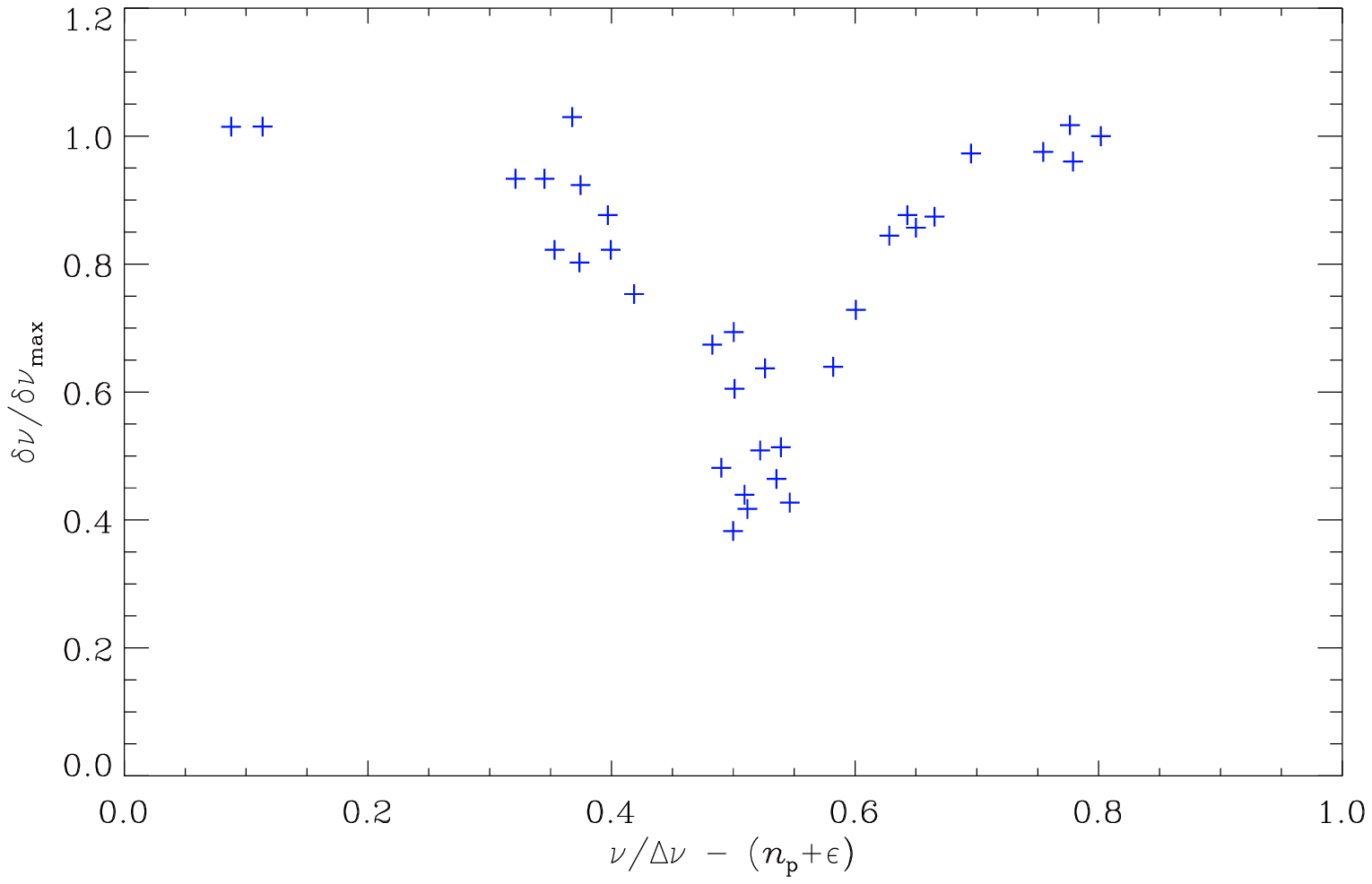}}
\caption{ {\it Top:} Observed rotational splittings  as a function of 
$\nu/\Delta \nu$ for the Kepler star KIC 5356201 (Beck et al. 2012). 
For this star, the mean large separation is  $\Delta \nu= 15.92 \mu$ Hz
 while  the mean g -mode spaging 
is $\Delta \Pi= 86.11$s and
the maximum  rotational splittings is  $\delta \nu_{\rm max}=0.405 $nHz.
 {\it Bottom:} Rotational splittings normalized to $\delta \nu_{\rm max}$ are 
   folded with   $(\nu-\nu_0)/\Delta \nu$. }
\label{fig:obs1}
\end{figure}

\section{The illustrative case of the red giant KIC 5356201}
\label{Obs}

Differential rotation of the red giant KIC 5356201 observed by \emph{Kepler} has been analyzed by 
\cite{2012Natur.481...55B}. As the inclination of the star has an intermediate value, 
the three components of the dipole mixed-mode multiplets are clearly visible, 
that allows a precise determination of the rotational splittings. 
With $\numax =  209.7 \pm 0.7 \mu$Hz, $\Dnu = 15.92 \pm 0.02\,\mu$Hz and
   a period spacing $\Delta \Pi= 86.2 \pm  0.03$ s, determined according to the method presented 
   in \cite{2012A&A...540A.143M}, the star lies at the bottom of the RGB, with 
   a seismically inferred radius of about $4.5\,R_\odot $. 
   The minimum rotational splitting given by \cite{2012Natur.481...55B} is 
   0.154 $\pm $0.003 $\mu$Hz; 
the authors  also mention that the average contrast between the maximum and minimum value of the splittings 
    is 1.7; the maximum value inferred by  \cite{2012A&A...540A.143M} is 
    0.405$\pm$0.010 $\mu$Hz. 
    With such global seismic parameters, one infers that this star is slightly more evolved 
    than model M1 but can be approximately represented by this model.

The observational rotational splittings are displayed as a function of $\nu/\Delta \nu$ 
($\nu$ is the frequency of the centroid mode of each identified multiplet) in Fig.~\ref{fig:obs1} (top).
The folded splittings are shown in Fig.~\ref{fig:obs1} (bottom). The global behavior of the curve 
 shows the same characteristics as  
 the theoretical equivalent in Fig.~\ref{fold1a} for model M1. Only  the value of the minimum
at $(\nu-\nu_0)/\Delta \nu$ is smaller than  0.4 (Fig.~\ref{fig:obs1}) 
for the observations compared with  0.59 for  model M1 (Fig.\ref{fold1a}).
 The difference is due to different values of the parameters
$\alpha_0$ ($\alpha_0=1.37 ~10^{-3}$  for the Kepler star).

Using the observed values  for $\numax, \Dnu$ and $\boldsymbol{\Delta} \Pi $ for the star KIC 5356201, 
we find $\eta= 0.49$ (Eq.~(\ref{eq:ratio6})) when 
 evaluated for the mode with frequency  $\nu/\Delta \nu=13.7$($\sim \nu\ind{max}/\Delta \nu$). This is close to the 
observed contrast  $\delta \nu/\delta \nu_{max}= 0.43$ 
 evaluated for the same mode.
   Note that if one corrects $\eta$ with the factor $f$ mentioned above, taking a typical 
   $f=0.8$ as for model M1,  one obtains
 $\eta= 0.45$  even  closer to the observed value. 

\begin{figure}[t]
\includegraphics[height=9cm,width=9.5cm]{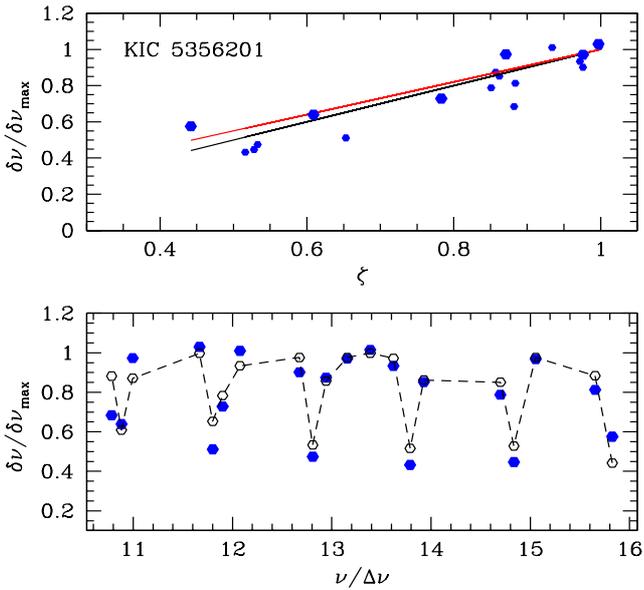}
\caption{{\it Top:} Observed rotational splittings as a function of 
$\zeta$  computed  with Eq.\ref{eq:zeta_app} and Eq.\ref{eq:chi10} 
for the Kepler star KIC 5358201 (blue dots). 
The solid curves are 
$\zeta (1-2 {\cal R}) + 2 {\cal R}$ with ${\cal R}=0$ (black) and $2 {\cal R}=0.1$ (red).  
{\it Bottom:} Observed rotational splittings (blue filled dots)as a function of $\nu/\Delta \nu$.
Theoretical rotational
splittings computed according to Eq.\ref{eq:split10} and Eq.\ref{eq:chi10} 
(black open dots connected with a
dashed line) with  ${\cal R}=0$.}
\label{fig:obs2}
\end{figure}

In Fig.~\ref{fig:obs2} (top), the observed rotational splittings normalized to their maximum value $\delta
\nu\ind{max}$ are plotted as a function of $\zeta$ obtained using Eqs.~(\ref{eq:zeta_app}) and 
 (\ref{eq:chi10}). The linear dependence of the
observed rotational splittings with $\zeta$ is clear.   
One single parameter, $2{\cal R}$, enters 
the expression for $\delta \nu/ \delta \nu_{max}$ (Eq.\ref{eq:ratio}) and
 can then be estimated to range between 
0. and 0.1 from Fig.~\ref{fig:obs2} (top). This indicates that for this star the ratio between 
the average core rotation and that of the convective envelope is larger than  20. 

We next compare the theoretical rotational splittings computed using Eqs.~(\ref{eq:chi10}) and (\ref{eq:split10}) with the  observed rotational splittings in Fig.~\ref{fig:obs2} (bottom).
 The agreement is quite acceptable and 
 validates our theoretical derivations of Sect.~\ref{interpretation}.

\section{Conclusion}
\label{conclusion}

We have investigated the properties of the rotational splittings of  dipole ($\ell=1$) modes of 
red giant models. 
The rotational splittings are computed with a  first-order perturbation method  because we focused on 
 slowly-rotating red giant stars. We considered two models  in  two different  H shell-burning evolutionary stages.  
We find that such modes  are either g-m modes (inertia and rotational splittings are fully dominated by the core properties) or p-m  modes (inertia is almost equally  shared between the core and the  envelope). 
For g-m modes, the  rotational splittings are  fully dominated by the average core rotation. As the time spent  by all considered g-m modes in the core is roughly the same, the resulting  mean core rotation corresponds to the rotation in the central layers weighted by the time spent by the modes in the core. For a radiative core, if the rotation is decreasing sharply from the center, the mean core rotation is  a lower limit of the central rotation. For p-m modes, the rotational splittings  remain  dominated by the core rotation but the contribution
of the envelope is no longer negligible. 

We find that the rotational splittings normalized to their maximum value linearly depend 
on the core contribution of mode inertia. The slope provides the ratio of the average envelope rotation to
the core one (${\cal R}$).  Mode inertia is univocally related to the  frequency difference between observed $\ell=1$ modes and the frequency of the closest radial mode. As this last quantity is measurable,   
 one can use the simple relation between  the  core inertia and this frequency difference to determine
the core inertia. This knowledge together with a measure of the rotational splitting then provides a  measure of ${\cal R}$.

As a step further in the modeling, we used asymptotic properties to show that 
the core contribution to  mode inertia relative to the total mode inertia depends on the frequency normalized to the large separation through a Lorentzian. 
This provides a theoretical support for the use of a Lorentzian profile for measuring the  observed splittings for
red giant stars \citep{2012arXiv1209.3336M}. 
This also led us to find  that the behavior of the rotational splittings of $\ell=1$ modes with the frequency for 
 slowly rotating core red giant stars depends  on three parameters only.  One is the large separation $\Delta \nu$,  the second is the g-mode period spacing $\Delta \Pi$, both can be determined observationally and can 
therefore provide again a measure of the third one,  ${\cal R}$. 

The contrast between the minimum $(\delta \nu_{\rm min})$ and maximum ($\delta \nu_{\rm max})$ values of
 the splittings can be evaluated at the frequency of maximum power. It  depends only on the above 3 
 parameters. For a core rotating much faster than the envelope (negligible ${\cal R}$), we obtain a 
 relation between $(\delta \nu_{\rm min})$, ($\delta \nu_{\rm max})$, $\Delta \nu$ and
$\Delta \Pi$. This can provide ($\delta \nu_{\rm max}$) (hence the average  core rotation) in case when the g-m modes have too small amplitudes to be detected and only $(\delta \nu_{\rm min})$ is available. 

 Based on the above theoretical developments,  we find that the Kepler red giant star 
  KIC 53656201  observed by Kepler (Beck et al. 2012)
   rotates with  a  rotation in the convective envelope 
  $\Omega_{CZ} < 0.05 ~\Omegacore \leq \Omega(r=0)$ so that the
  core is rotating more than 20 times faster than the envelope.

 The present understanding of the properties of  rotational splittings  is a contribution
to the effort to decipher the effect of slow  rotation  on red giant star frequency spectra 
 and establish seismic diagnostics on rotation and transport of angular momentum. 
 For more rapidly rotating red giant stars, non-perturbative methods  must be used. This will be the subject of the third paper of this series. 
 
 \begin{acknowledgements}
 JPM acknowledges financial support through a 3 year CDD contract with CNES. 
R-M.O. is indebted to the "F\'ed\'eration Wallonie- Bruxelles
- Fonds Sp\'eciaux pour la Recherche / Cr\'edit de d\'emarrage - Universit\'e de Li\`ege" for financial
support. The authors also acknowledge financial
support from the French National Research Agency (ANR) for the project ANR-07-BLAN-0226 SIROCO (SeIsmology, ROtation and COnvection with the CoRoT satellite). We also thank the anonymous referee  for a careful reading of the manuscript and useful suggestions.
\end{acknowledgements}


\begin{thebibliography}{39}
\expandafter\ifx\csname natexlab\endcsname\relax\def\natexlab#1{#1}\fi

\bibitem[{{Baglin} {et~al.}(2006){Baglin}, {Auvergne}, {Boisnard}, {Lam-Trong},
  {Barge}, {Catala}, {Deleuil}, {Michel}, \& {Weiss}}]{2006cosp...36.3749B}
{Baglin}, A., {Auvergne}, M., {Boisnard}, L., {et~al.} 2006, in 36th COSPAR
  Scientific Assembly, Vol.~36, 3749

\bibitem[{{Ballot} {et~al.}(2011){Ballot}, {Ligni{\`e}res}, {Prat}, {Reese}, \&
  {Rieutord}}]{2011arXiv1109.6856B}
{Ballot}, J., {Ligni{\`e}res}, F., {Prat}, V., {Reese}, D.~R., \& {Rieutord},
  M. 2011, ArXiv e-prints

\bibitem[{{Ballot} {et~al.}(2010){Ballot}, {Ligni{\`e}res}, {Reese}, \&
  {Rieutord}}]{2010A&A...518A..30B}
{Ballot}, J., {Ligni{\`e}res}, F., {Reese}, D.~R., \& {Rieutord}, M. 2010,
  \aap, 518, A30

\bibitem[{{Beck} {et~al.}(2012){Beck}, {Montalban}, {Kallinger}, {De Ridder},
  {Aerts}, {Garc{\'{\i}}a}, {Hekker}, {Dupret}, {Mosser}, {Eggenberger},
  {Stello}, {Elsworth}, {Frandsen}, {Carrier}, {Hillen}, {Gruberbauer},
  {Christensen-Dalsgaard}, {Miglio}, {Valentini}, {Bedding}, {Kjeldsen},
  {Girouard}, {Hall}, \& {Ibrahim}}]{2012Natur.481...55B}
{Beck}, P.~G., {Montalban}, J., {Kallinger}, T., {et~al.} 2012, \nat, 481, 55

\bibitem[{{Bedding} \& {Kjeldsen}(2003)}]{2003PASA...20..203B}
{Bedding}, T.~R. \& {Kjeldsen}, H. 2003, \pasa, 20, 203

\bibitem[{{Bedding} {et~al.}(2011){Bedding}, {Mosser}, {Huber},
  {Montalb{\'a}n}, {Beck}, {Christensen-Dalsgaard}, {Elsworth},
  {Garc{\'{\i}}a}, {Miglio}, {Stello}, {White}, {De Ridder}, {Hekker}, {Aerts},
  {Barban}, {Belkacem}, {Broomhall}, {Brown}, {Buzasi}, {Carrier}, {Chaplin},
  {di Mauro}, {Dupret}, {Frandsen}, {Gilliland}, {Goupil}, {Jenkins},
  {Kallinger}, {Kawaler}, {Kjeldsen}, {Mathur}, {Noels}, {Aguirre}, \&
  {Ventura}}]{2011Natur.471..608B}
{Bedding}, T.~R., {Mosser}, B., {Huber}, D., {et~al.} 2011, \nat, 471, 608

\bibitem[{{Belkacem} {et~al.}(2011){Belkacem}, {Goupil}, {Dupret}, {Samadi},
  {Baudin}, {Noels}, \& {Mosser}}]{2011A&A...530A.142B}
{Belkacem}, K., {Goupil}, M.~J., {Dupret}, M.~A., {et~al.} 2011, \aap, 530,
  A142

\bibitem[{{Borucki} {et~al.}(2010){Borucki}, {Koch}, {Basri}, {Batalha},
  {Brown}, {Caldwell}, {Caldwell}, {Christensen-Dalsgaard}, {Cochran},
  {DeVore}, {Dunham}, {Dupree}, {Gautier}, {Geary}, {Gilliland}, {Gould},
  {Howell}, {Jenkins}, {Kondo}, {Latham}, {Marcy}, {Meibom}, {Kjeldsen},
  {Lissauer}, {Monet}, {Morrison}, {Sasselov}, {Tarter}, {Boss}, {Brownlee},
  {Owen}, {Buzasi}, {Charbonneau}, {Doyle}, {Fortney}, {Ford}, {Holman},
  {Seager}, {Steffen}, {Welsh}, {Rowe}, {Anderson}, {Buchhave}, {Ciardi},
  {Walkowicz}, {Sherry}, {Horch}, {Isaacson}, {Everett}, {Fischer}, {Torres},
  {Johnson}, {Endl}, {MacQueen}, {Bryson}, {Dotson}, {Haas}, {Kolodziejczak},
  {Van Cleve}, {Chandrasekaran}, {Twicken}, {Quintana}, {Clarke}, {Allen},
  {Li}, {Wu}, {Tenenbaum}, {Verner}, {Bruhweiler}, {Barnes}, \&
  {Prsa}}]{2010Sci...327..977B}
{Borucki}, W.~J., {Koch}, D., {Basri}, G., {et~al.} 2010, Science, 327, 977

\bibitem[{{Brown} {et~al.}(1991){Brown}, {Gilliland}, {Noyes}, \&
  {Ramsey}}]{1991ApJ...368..599B}
{Brown}, T.~M., {Gilliland}, R.~L., {Noyes}, R.~W., \& {Ramsey}, L.~W. 1991,
  \apj, 368, 599

\bibitem[{{Christensen-Dalsgaard}(2008)}]{2008Ap&SS.316..113C}
{Christensen-Dalsgaard}, J. 2008, \apss, 316, 113

\bibitem[{{Christensen-Dalsgaard} \& {Berthomieu}(1991)}]{1991sia..book..401C}
{Christensen-Dalsgaard}, J. \& {Berthomieu}, G. 1991, {Theory of solar
  oscillations}, ed. A.~N. {Cox}, W.~C. {Livingston}, \& M.~S. {Matthews},
  401--478

\bibitem[{{De Ridder} {et~al.}(2009){De Ridder}, {Barban}, {Baudin}, {Carrier},
  {Hatzes}, {Hekker}, {Kallinger}, {Weiss}, {Baglin}, {Auvergne}, {Samadi},
  {Barge}, \& {Deleuil}}]{2009Natur.459..398D}
{De Ridder}, J., {Barban}, C., {Baudin}, F., {et~al.} 2009, \nat, 459, 398

\bibitem[{{Deheuvels} {et~al.}(2012){Deheuvels}, {Garc{\'{\i}}a}, {Chaplin},
  {Basu}, {Antia}, {Appourchaux}, {Benomar}, {Davies}, {Elsworth}, {Gizon},
  {Goupil}, {Reese}, {Regulo}, {Schou}, {Stahn}, {Casagrande},
  {Christensen-Dalsgaard}, {Fischer}, {Hekker}, {Kjeldsen}, {Mathur}, {Mosser},
  {Pinsonneault}, {Valenti}, {Christiansen}, {Kinemuchi}, \&
  {Mullally}}]{2012ApJ...756...19D}
{Deheuvels}, S., {Garc{\'{\i}}a}, R.~A., {Chaplin}, W.~J., {et~al.} 2012, \apj,
  756, 19

\bibitem[{{Dupret} {et~al.}(2009){Dupret}, {Belkacem}, {Samadi}, {Montalban},
  {Moreira}, {Miglio}, {Godart}, {Ventura}, {Ludwig}, {Grigahc{\`e}ne},
  {Goupil}, {Noels}, \& {Caffau}}]{2009A&A...506...57D}
{Dupret}, M.-A., {Belkacem}, K., {Samadi}, R., {et~al.} 2009, \aap, 506, 57

\bibitem[{{Dziembowski}(1977)}]{1977AcA....27...95D}
{Dziembowski}, W. 1977, \actaa, 27, 95

\bibitem[{{Dziembowski}(1971)}]{1971AcA....21..289D}
{Dziembowski}, W.~A. 1971, \actaa, 21, 289

\bibitem[{{Dziembowski} {et~al.}(2001){Dziembowski}, {Gough}, {Houdek}, \&
  {Sienkiewicz}}]{2001MNRAS.328..601D}
{Dziembowski}, W.~A., {Gough}, D.~O., {Houdek}, G., \& {Sienkiewicz}, R. 2001,
  \mnras, 328, 601

\bibitem[{{Eggenberger} {et~al.}(2012){Eggenberger}, {Montalb{\'a}n}, \&
  {Miglio}}]{2012A&A...544L...4E}
{Eggenberger}, P., {Montalb{\'a}n}, J., \& {Miglio}, A. 2012, \aap, 544, L4

\bibitem[{{Hekker} {et~al.}(2009){Hekker}, {Kallinger}, {Baudin}, {De Ridder},
  {Barban}, {Carrier}, {Hatzes}, {Weiss}, \& {Baglin}}]{2009A&A...506..465H}
{Hekker}, S., {Kallinger}, T., {Baudin}, F., {et~al.} 2009, \aap, 506, 465

\bibitem[{{Kjeldsen} \& {Bedding}(1995)}]{1995A&A...293...87K}
{Kjeldsen}, H. \& {Bedding}, T.~R. 1995, \aap, 293, 87

\bibitem[{{Ledoux}(1951)}]{1951ApJ...114..373L}
{Ledoux}, P. 1951, \apj, 114, 373

\bibitem[{{Ligni{\`e}res}(2011)}]{2011LNP...832..259L}
{Ligni{\`e}res}, F. 2011, in Lecture Notes in Physics, Berlin Springer Verlag,
  Vol. 832, Lecture Notes in Physics, Berlin Springer Verlag, ed. J.-P.
  {Rozelot} \& C.~{Neiner}, 259

\bibitem[{{Marques} {et~al.}(2012){Marques}, {Goupil}, {Lebreton}, {Talon},
  {Palacios}, {Ouazzani}, {Mosser}, {Moya}, {Morel}, {Pichon}, {Mathis},
  {Zahn}, {Nghiem}, \& {Turck-Chi{\`e}ze}}]{Marques2012}
{Marques}, J.~P., {Goupil}, M.~J., {Lebreton}, Y., {et~al.} 2012, \aap
  (submitted)

\bibitem[{{Meynet} {et~al.}(2012){Meynet}, {Ekstrom}, {Maeder}, {Eggenberger},
  {Saio}, {Chomienne}, \& {Haemmerl{\'e}}}]{Meynet2012}
{Meynet}, G., {Ekstrom}, S., {Maeder}, A., {et~al.} 2012, Lectures Notes in
  Physics (in press)

\bibitem[{{Montalb{\'a}n} {et~al.}(2010){Montalb{\'a}n}, {Miglio}, {Noels},
  {Scuflaire}, \& {Ventura}}]{2010ApJ...721L.182M}
{Montalb{\'a}n}, J., {Miglio}, A., {Noels}, A., {Scuflaire}, R., \& {Ventura},
  P. 2010, \apjl, 721, L182

\bibitem[{{Mosser} {et~al.}(2011{\natexlab{a}}){Mosser}, {Barban},
  {Montalb{\'a}n}, {Beck}, {Miglio}, {Belkacem}, {Goupil}, {Hekker}, {De
  Ridder}, {Dupret}, {Elsworth}, {Noels}, {Baudin}, {Michel}, {Samadi},
  {Auvergne}, {Baglin}, \& {Catala}}]{2011A&A...532A..86M}
{Mosser}, B., {Barban}, C., {Montalb{\'a}n}, J., {et~al.} 2011{\natexlab{a}},
  \aap, 532, A86

\bibitem[{{Mosser} {et~al.}(2011{\natexlab{b}}){Mosser}, {Belkacem}, {Goupil},
  {Michel}, {Elsworth}, {Barban}, {Kallinger}, {Hekker}, {De Ridder}, {Samadi},
  {Baudin}, {Pinheiro}, {Auvergne}, {Baglin}, \&
  {Catala}}]{2011A&A...525L...9M}
{Mosser}, B., {Belkacem}, K., {Goupil}, M.~J., {et~al.} 2011{\natexlab{b}},
  \aap, 525, L9

\bibitem[{{Mosser} {et~al.}(2010){Mosser}, {Belkacem}, {Goupil}, {Miglio},
  {Morel}, {Barban}, {Baudin}, {Hekker}, {Samadi}, {De Ridder}, {Weiss},
  {Auvergne}, \& {Baglin}}]{2010A&A...517A..22M}
{Mosser}, B., {Belkacem}, K., {Goupil}, M.-J., {et~al.} 2010, \aap, 517, A22

\bibitem[{{Mosser} {et~al.}(2012{\natexlab{a}}){Mosser}, {Goupil}, {Belkacem},
  {Marques}, {Beck}, {Bloemen}, {De Ridder}, {Barban}, {Deheuvels}, {Elsworth},
  {Hekker}, {Kallinger}, {Ouazzani}, {Pinsonneault}, {Samadi}, {Stello},
  {Garcia}, {Klaus}, {Li}, {Mathur}, \& {Morris}}]{2012arXiv1209.3336M}
{Mosser}, B., {Goupil}, M.~J., {Belkacem}, K., {et~al.} 2012{\natexlab{a}},
  ArXiv e-prints 1209.3336

\bibitem[{{Mosser} {et~al.}(2012{\natexlab{b}}){Mosser}, {Goupil}, {Belkacem},
  {Michel}, {Stello}, {Marques}, {Elsworth}, {Barban}, {Beck}, {Bedding}, {De
  Ridder}, {Garc{\'{\i}}a}, {Hekker}, {Kallinger}, {Samadi}, {Stumpe},
  {Barclay}, \& {Burke}}]{2012A&A...540A.143M}
{Mosser}, B., {Goupil}, M.~J., {Belkacem}, K., {et~al.} 2012{\natexlab{b}},
  \aap, 540, A143

\bibitem[{{Mosser} {et~al.}(2013){Mosser}, {Michel}, {Belkacem}, {Goupil},
  {Baglin}, {Barban}, {Provost}, {Samadi}, {Auvergne}, \& {Catala}}]{mesure}
{Mosser}, B., {Michel}, E., {Belkacem}, K., {et~al.} 2013, submitted to A\&A

\bibitem[{{Osaki}(1975)}]{1975PASJ...27..237O}
{Osaki}, Y. 1975, \pasj, 27, 237

\bibitem[{{Ouazzani} {et~al.}(2012){Ouazzani}, {Dupret}, \&
  {Reese}}]{2012arXiv1209.5621O}
{Ouazzani}, R.-M., {Dupret}, M.-A., \& {Reese}, D. 2012, ArXiv e-prints

\bibitem[{{Reese} {et~al.}(2006){Reese}, {Ligni{\`e}res}, \&
  {Rieutord}}]{2006A&A...455..621R}
{Reese}, D., {Ligni{\`e}res}, F., \& {Rieutord}, M. 2006, \aap, 455, 621

\bibitem[{{Samadi} {et~al.}(2012){Samadi}, {Belkacem}, {Dupret}, {Ludwig},
  {Baudin}, {Caffau}, {Goupil}, \& {Barban}}]{2012A&A...543A.120S}
{Samadi}, R., {Belkacem}, K., {Dupret}, M.-A., {et~al.} 2012, \aap, 543, A120

\bibitem[{{Scuflaire}(1974)}]{1974A&A....36..107S}
{Scuflaire}, R. 1974, \aap, 36, 107

\bibitem[{{Shibahashi}(1979)}]{1979PASJ...31...87S}
{Shibahashi}, H. 1979, \pasj, 31, 87

\bibitem[{{Tassoul}(1980)}]{1980ApJS...43..469T}
{Tassoul}, M. 1980, \apjs, 43, 469

\bibitem[{{Unno} {et~al.}(1989){Unno}, {Osaki}, {Ando}, {Saio}, \&
  {Shibahashi}}]{Unno89}
{Unno}, W., {Osaki}, Y., {Ando}, H., {Saio}, H., \& {Shibahashi}, H. 1989,
  {Nonradial oscillations of stars, Tokyo: University of Tokyo Press, 1989, 2nd
  ed.}

\end{thebibliography}

\appendix

\section{Properties of eigenfunctions and mode inertia } 

\subsection{Displacement eigenfunctions}
\label{eigenf}
\begin{figure}[t]
\centering
\includegraphics[height=9cm,width=9.5cm]{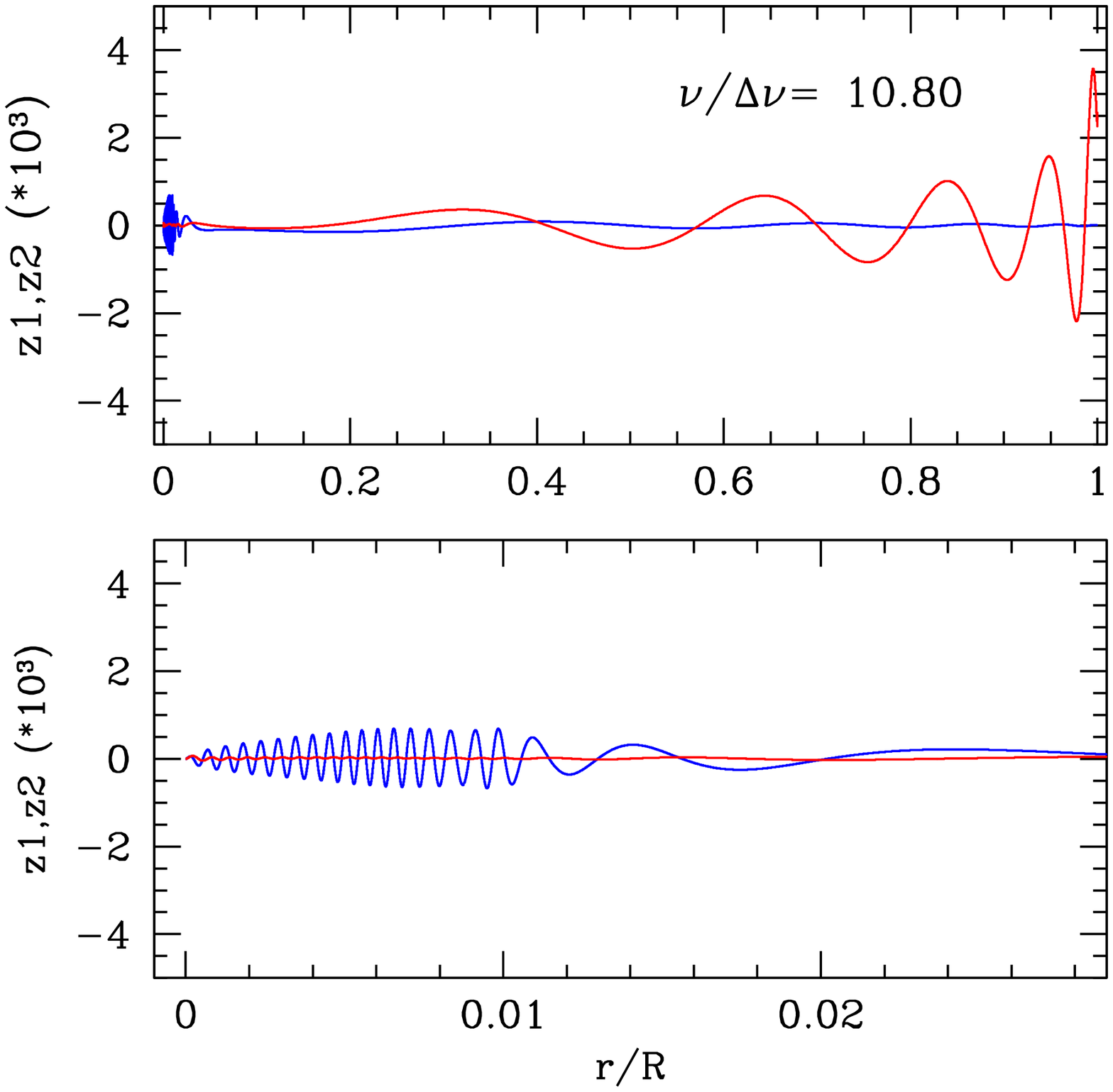}
\includegraphics[height=9cm,width=9.5cm]{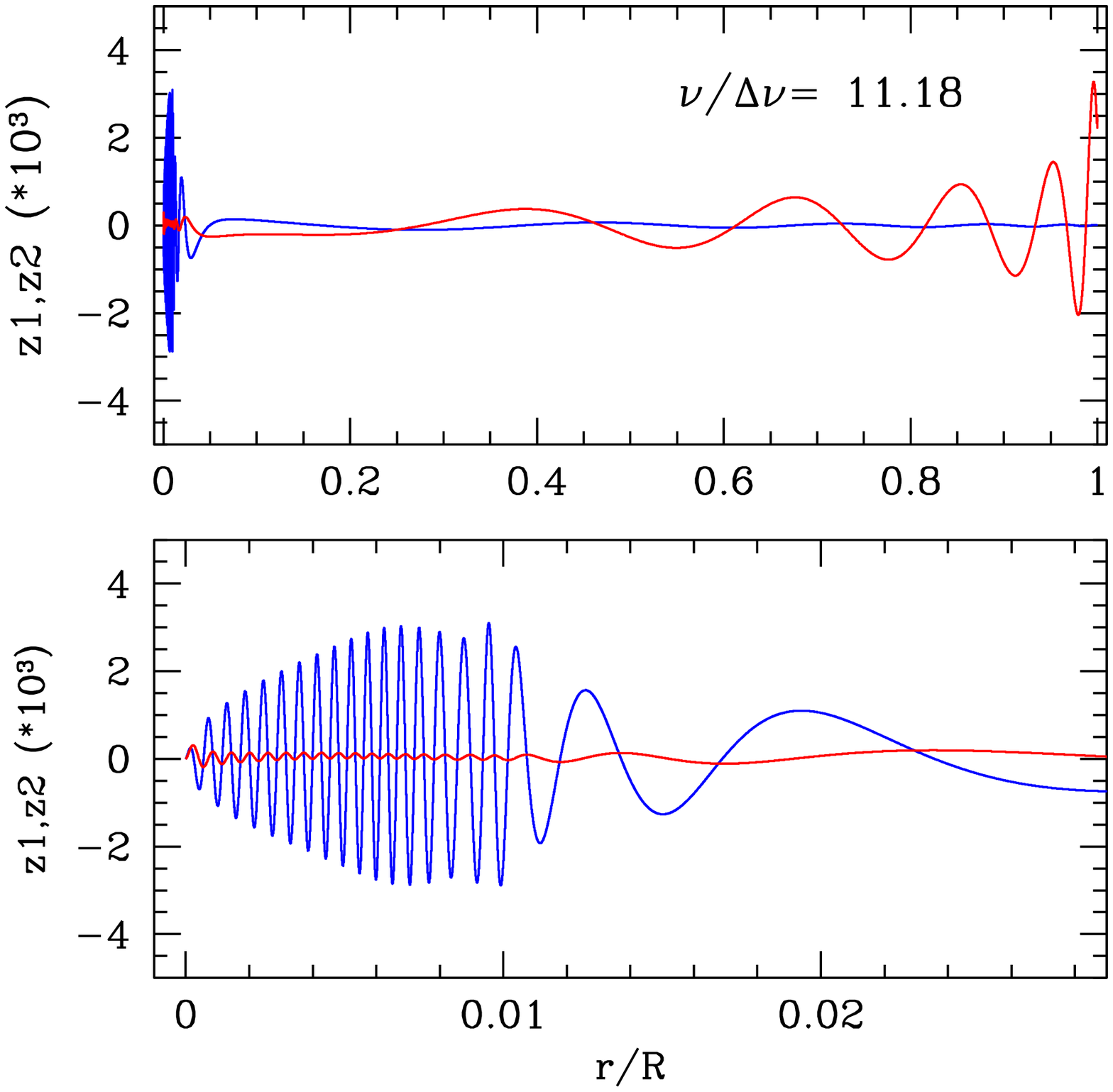}
\caption{\label{eigenf} Displacement eigenfunctions as a function of the normalized
radius $r/R$, in blue the horizontal  component to the displacement eigenfunction $z_2$ (Eq.~\ref{newvar2}) and
 in red the radial  component to the  displacement eigenfunction $z_1$ (Eq.~\ref{newvar1}). 
From top to bottom  $\ell=1$ modes  1 and 3 of Table.~\ref{Tab2}.
 }
\label{zb}
\end{figure}

The displacement eigenfunctions computed  for two modes of the selected  pattern (see Table~\ref{Tab2}) for model M1 are shown as a function of the normalized radius $r/R$ in Fig.~\ref{zb}. They correspond to the modes with the largest and smallest maximum  amplitudes in the pattern.  
 The  inner part is dominated by the horizontal displacement $z_2$ and oscillates with
  a large number of nodes, typical of a high-order gravity-mode. The largest maximum 
  amplitude corresponds to the most g-dominated mode 
   whereas the smallest maximum amplitudes  arise for the p-m  modes $\nu_1$ and $\nu_6$. 

The maximum amplitude of $z_2$ occurs deep in the
 g-cavity, at the same radius for all modes of the pattern. The region of non-negligible
 amplitude defines the radius of  a \emph{seismic rotating core} which 
 is found  here independent of the mode  ($r/R \sim 0.02$) and far smaller than the  upper turning
radius  of the inner gravity resonant cavity ($x_2 \sim 0.08$, Table~\ref{Tab3})

\subsection{Behavior of  $\beta$ and $\beta\ind{core}$ with $\zeta$ }
\label{beta}

For $\ell=1$ modes of red giants, the term $z_2z_1$ in $\beta$  (Eq.~(\ref{eq:beta})) plays almost no role 
because $z_2 z_1 \ll z_2^2$ in the core and $z_2 z_1  \ll z_1^2$ in the envelope (see Fig.~\ref{zb}). As a
result, we have
 \begin{eqnarray}
 \beta &\approx& \frac{1}{I}~\int_0^1  ~\left(z_1^2+ z_2^2- \frac{1}{2}~z_2^2\right)\, ~ \frac{{\rm d}x}{x}\,  \\
 &=& 1-\frac{1}{2}~\frac{1}{I}~\int_0^1 ~z_2^2\, ~ \frac{{\rm d}x}{x}\, \\ 
 &=& 1-\frac{1}{2} \zeta \, .
\label{eq:beta2}
\end{eqnarray}
where $\zeta$ is defined in Eq.\ref{eq:zeta}.
The linear dependence of $\beta$ with $\zeta$  is verified in Fig.~\ref{betacore}. Furthermore,
for all modes,  $z_2^2 \gg  z_1^2$, $2z_1 z_2$ in the g- cavity (see Fig.\ref{zb})
then   $\beta\ind{core,nl} \approx \beta\ind{core}$ where for $l=1$ modes,
 we derive 
 \begin{eqnarray}
 \beta\ind{core} &\approx& \frac{1}{I}~\int\ind{core}  ~\left(z_1^2+ z_2^2- \frac{1}{2}~z_2^2\right)\, ~ \frac{{\rm d}x}{x}  \\
 &\approx& \frac{1}{2I}~\int\ind{core} ~z_2^2\, ~ \frac{{\rm d}x}{x} =\frac{1}{2}\, \frac{\Icore}{I}=\frac{1}{2} \; \zeta \, ,
\label{eq:betacore1}
\end{eqnarray}
hence
 \begin{eqnarray}
 \beta\ind{env} &=& \beta- \beta\ind{core} \approx 1-\zeta \, .
\label{eq:betaenv}
\end{eqnarray}

\begin{figure}[t]
\includegraphics[height=9cm,width=9.5cm]{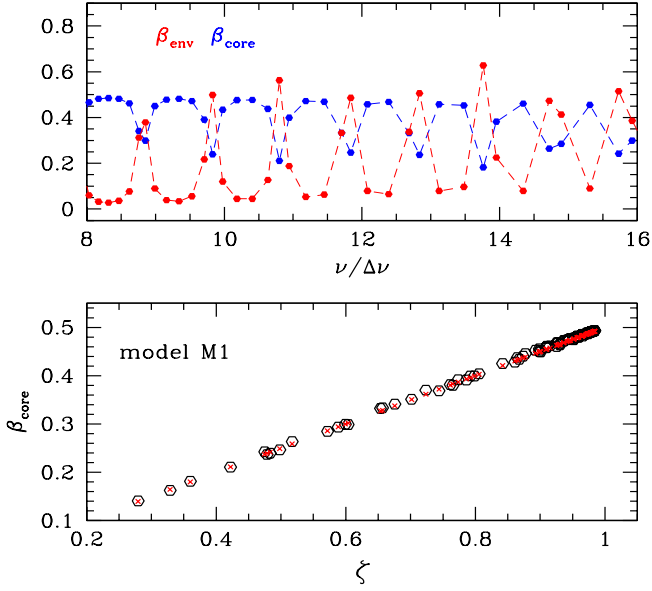}
\caption{{\it Top:} $\beta\ind{env}$ and $\beta\ind{core}$ as a function of $\nu/\Delta \nu$
for model M1. {\it Bottom} Same as top for the ratio  $\beta\ind{core}$  
 as a function of $\zeta$ (black open dots). The
approximation $\beta=1-(1/2)~\zeta$  using the numerical values of $\zeta$ is represented with red crosses.}
\label{betacore} 
\end{figure}

Numerical values for model M1 confirm that $\beta\ind{core}$  increases linearly with
 $\zeta$ with a slope 1/2 (Fig.~\ref{betacore}).
For g-m modes ($\zeta\sim 1$), $\beta\ind{core}$ dominates  with a nearly constant value of 0.5. 
P-m modes correspond to the \emph{teeth} of the saw type variation of $\beta_{env}$ and the
ratio $\beta\ind{env}/\beta\ind{core}\sim 0.25 $ (Fig.~\ref{betacore}).

\subsection{An approximate expression for   $\zeta$}
\label{zeta}

This section determines an approximate expression of  $\zeta=\Icore/I$ as a function of 
$\nu/\Delta \nu$. The derivation is based on results of an asymptotic method developed by
\cite{1979PASJ...31...87S} to which we refer for details \citep[see also][]{Unno89}.

\begin{itemize}

\item The envelope ($\sim$p-propagative cavity) is characterized by  $z_2^2 \ll z_1^2$. 
Using Eq. (16.47) from Unno et al., it is straightforward to derive the following 
approximate expression
 \begin{eqnarray}\label{z12}
z_1^2 ~\frac{{\rm d}x}{x}&\sim & \frac{c^2}{\sigma} ~ \Bigl(1-\sin\sigma \tau\Bigr)~{\rm d}\tau \, .
 \end{eqnarray}
The constant $c$ can be  determined by the condition $\xir=1$ at the surface.  
 \begin{eqnarray}
\tau(x_3,x) &=& \frac{2}{\sigma}~ \int_{x_3}^x ~k\ind{r} ~{\rm d}x'  \, , 
 \end{eqnarray}
 with 
 \begin{eqnarray}\label{krp}
k\ind{r} &\sim& \frac{1}{c_s} ~\Bigl(\sigma^2-S_l^2\Bigr)^{1/2}\, .
 \end{eqnarray}
 where we have assumed $\sigma^2>>N^2$ in Eq.\ref{kr2true}. 
 Note that in the process of deriving the 
  amplitude of $z_1^2$ arising in front of the sinusoidal term in
Eq.\ref{z12},   one can neglect $S_l^2$ in front of $\sigma^2$ 
in the expression for $k\ind{r}$ (i.e. $k\ind{r} \sim \sigma/c_s$). However  this is not the case 
when  $k\ind{r}$ is  in the phase of the sinusoidal term where we keep the expression 
Eq.\ref{krp}.

The inertia in the envelope can then be approximated as: 
 \begin{eqnarray}
 \label{Ienv}
I\ind{env} &\sim& \int\ind{env}~ z_1^2 ~\frac{{\rm d}x}{x}  \label{Ienv} \\
&\sim&  \frac{c^2}{\sigma} \tau_p ~
~\left(1-\frac{\cos(\sigma \tau_p)-1}{\sigma\tau_p}\right)
\sim  \frac{c^2}{\sigma} ~\tau_p \, ,  \label{Ienv2}
 \end{eqnarray}
where we have defined 
 \begin{eqnarray}
 \label{eq:taup}
\tau_p &=& \frac{2}{\sigma}~ \int_{x_3}^{1} ~\Bigl(\sigma^2-S_l^2\Bigr)^{1/2} \frac{{\rm d}x}{c_s}
= \frac{2\pi}{ \sigma}~ \frac{1}{ f}  ~\frac{\nu}{ \Delta \nu} \, ,
 \end{eqnarray}
and  the mean large separation is
 \begin{eqnarray}
\Delta \nu \equiv \left(2~\int_{0}^{1} ~  \frac{{\rm d}x}{c_s}\right)^{-1} \, .
 \end{eqnarray}
 The factor $f$ is of  order unity and represents the  
 difference  between  the integration
 from $x_3$ and from the center. We take $f=1$ unless specified otherwise. 
 The last equality in Eq.~(\ref{Ienv2}) is obtained assuming $\sigma \tau_p\gg 1$. 

\medskip 

\item The core ($\sim$g-propagative cavity) is characterized by $z_2^2 \gg  z_1^2$. Again, the asymptotic
results lead to the following expression
 \begin{eqnarray}
z_2^2 ~\frac{{\rm d}x}{x} &\sim&  \frac{ a^2}{\sigma} ~\Bigl(1-\sin \sigma \tau\Bigr) ~  {\rm d}\tau \, , 
\end{eqnarray}
where $a$ is a constant that  is  determined by the resonant frequency condition  between 
the p and g cavities and 
 \begin{eqnarray}
\tau (x_1,x) &= &  \frac{2}{\sigma}~ \int\ind{x_1}^x ~k\ind{r} ~{\rm d}x' \, , 
\end{eqnarray}
and we have used
 \begin{eqnarray}
k\ind{r} &\sim& \frac{\sqrt{\Lambda}}{\sigma x} ~ (N^2-\sigma^2)^{1/2} \, .
\end{eqnarray}

Hence recalling that $\sigma \tau_g \gg 1$, the inertia in the core can be approximated as 
 \begin{eqnarray}
\Icore&\sim& \frac{a^2}{\sigma}  ~\tau_g~
\left(1+ \frac{\cos(\sigma \tau_g )-1}{\sigma \tau_g}\right) \sim \frac{a^2}{\sigma} ~\tau_g \, , 
\end{eqnarray}
where we have defined
 \begin{eqnarray}
 \label{eq:taug}
 \tau_g &=&  \tau(x1,x2)= 
\frac{\sqrt{\Lambda}}{\sigma^2}~ \int\ind{core} ~ (N^2-\sigma^2)^{1/2}~ \frac{{\rm d}x}{x} \, .
 \end{eqnarray}

\medskip 

\item The  ratio $q \equiv I\ind{env}/\Icore$  is then approximated  by
 \begin{eqnarray}
  q &\approx & \left(\frac{c}{a}\right)^2 ~\frac{\tau_p}{\tau_g} \, .
  \end{eqnarray}
  
\medskip

\item   We  obtain the ratio $c/a$ (from Unno et al's Eq. (16.49) and Eq. (16.50)) as 
 \begin{eqnarray}\label{csa}
\frac{c}{a}&=& 2\; \frac{\cos(\sigma \tau_g/2)}{\cos(\sigma \tau_p/2)} 
\sim \pm 2 ~ \cos\left(\frac{\sigma \tau_g}{2}\right) \, , 
  \end{eqnarray}
where we have used the fact that $\sigma \tau_p \sim 2 n_p \pi$. 
A exponential term is present in Unno et
al's expression  with the argument being an integral over the evanescent region between the p- and
g-cavities.
 As this region is quite narrow in our models for  the considered modes, 
 the exponential is taken to be 1. Nevertheless the width of the 
 evanescent region  depends on the considered mode and in some cases, 
  for accurate quantitative results, it might be necessary to include
 effects of the evanescent zone with a finite width.

\item The  ratio $q$  is then eventually approximated  by 
 \begin{eqnarray}
 \label{eq:ratio2}
 q&= & 4~\cos^2\left(\frac{\sigma \tau_g}{2}\right) ~ \frac{\tau_p}{\tau_g} \, .
  \end{eqnarray}

\medskip 
\item  

For the relative core inertia $\zeta=\Icore/I$, 
 \begin{equation}
\zeta = \frac{1}{1+q} \approx \Bigl( 1+4~\cos^2\Bigl(\frac{\sigma \tau_g}{2}\Bigr) 
~ \frac{\tau_p}{\tau_g} \Bigr)^{-1} \, .
 \end{equation}

\medskip 

\item  We now use the approximate expressions Eq.~(\ref{eq:taup}) and Eq.~(\ref{eq:taug}) 
in order to  derive for the ratio $\tau_p/\tau_g$ in terms of observable quantities 
 \begin{eqnarray}
\label{taupstaug}
 \frac{\tau_p}{\tau_g} \approx  \frac{1} {f}~ \alpha_0~ y^2 \, , 
 \end{eqnarray}
where for convenience  we hase defined $y=\nu/\Delta \nu $ and
\begin{eqnarray} 
\alpha_0 &=& \Delta \nu ~ \Delta \Pi \, ,
 \end{eqnarray}
with the period spacing for g modes 
\begin{eqnarray} 
 \Delta \Pi &=& \frac{2 \pi^2}{\sqrt{\Lambda}}~ \left( \int\ind{core} 
 ~ (N^2-\sigma^2)^{1/2} \,\frac{{\rm d}x}{x} \right)^{-1} \, .
\end{eqnarray}

We also write 
 \begin{equation}
 \sigma \tau_g = ~ \frac{2 \pi} {\Delta \Pi  ~ \nu} \, , 
 \end{equation}
so that we obtain 
 \begin{eqnarray}
\zeta & \approx& \frac{1}{1+  \alpha_0 ~\chi^2/f}  \approx \frac{1}{1+  \alpha_0 ~\chi^2}  \\
\chi &=& 2~\frac{\nu}{\Delta \nu}~\cos\left(\frac{\pi} {\Delta \Pi  ~ \nu}\right) \, .
 \end{eqnarray}

\end{itemize}

\end{document}